\documentclass[aps,pra,superscriptaddress,floatfix,notitlepage,nofootinbib,longbibliography,11pt]{revtex4-1} 
\pdfoutput=1
\usepackage{amsmath,amsthm,amsfonts,amssymb,braket}
\usepackage{graphicx}
\usepackage{color}
\usepackage{newtxmath}

\numberwithin{equation}{section}
\numberwithin{figure}{section}

\makeatletter
\def\l@subsubsection#1#2{}
\makeatother

\usepackage[colorlinks=true,citecolor=blue,linkcolor=blue]{hyperref}
\usepackage[capitalize,nameinlink]{cleveref}

\newtheorem{lemma}{Lemma}[section]
\newtheorem{theorem}[lemma]{Theorem}

\theoremstyle{definition}

\newcommand{\CC}{{\mathbb{C}}}
\newcommand{\RR}{{\mathbb{R}}}
\newcommand{\ZZ}{{\mathbb{Z}}}
\newcommand{\Z}{{\mathbb{Z}}}
\newcommand{\FF}{{\mathbb{F}}}

\newcommand{\tX}{{\tilde{X}}}
\newcommand{\tZ}{{\tilde{Z}}}
\newcommand{\tG}{{\tilde{G}}}
\newcommand{\tU}{{\tilde{U}}}
\newcommand{\tbeta}{{\tilde{\beta}}}
\newcommand{\tdbeta}{{\tilde{d\beta}}}
\newcommand{\hsquare}{{\hat{\square}}}

\newcommand{\one}{{\mathbf{1}}}
\newcommand{\diag}{{\mathrm{diag}}}

\newcommand{\rd}{{\mathrm{d}}}

\newcommand{\projz}{{\mathbf{z}}}

\DeclareMathOperator{\Sq}{\mathrm{Sq}}
\DeclareMathOperator{\depth}{\mathrm{depth}}
\DeclareMathOperator{\Arf}{\mathrm{Arf}}

\begin{document}
\title{A QCA for every SPT}

\begin{abstract}
In three dimensions, there is a nontrivial quantum cellular automaton (QCA) 
which disentangles \cite{Haah_2022}  the three-fermion Walker--Wang model, 
a model whose action depends on Stiefel--Whitney classes of the spacetime manifold.
Here we present a conjectured generalization to higher dimensions.
For an arbitrary symmetry protected topological phase of time reversal 
whose action depends on Stiefel--Whitney classes, 
we construct a corresponding QCA that we conjecture disentangles that phase.  
Some of our QCA are Clifford, and we relate these to a classification theorem of Clifford QCA.
We also identify Clifford QCA in $4m+1$ dimensions, 
for which we find a low-depth circuit description using non-Clifford gates
but not with Clifford gates.
\end{abstract}

\author{Lukasz Fidkowski}
\affiliation{Department of Physics, University of Washington, Seattle WA USA}

\author{Jeongwan Haah}
\affiliation{Microsoft Quantum, Redmond, WA USA}

\author{Matthew B.~Hastings}
\affiliation{Microsoft Quantum, Redmond, WA USA}
\maketitle

\tableofcontents

\section{Introduction}

Quantum Cellular Automata (QCA) are locality-preserving operator algebra automorphisms on lattice spin systems.  
For finite size systems, 
one can view a QCA as being the conjugation action by a unitary operator, uniquely defined up to phase.%
\footnote{%
	In the following, 
	we will often not make a distinction between QCA and the corresponding unitaries.
}
After their introduction in the quantum information community~\cite{GNVW},
it was realized that QCA have applications in condensed matter physics, 
for example in the classification of Floquet many-body localized (MBL) 
phases~\cite{Po_2016, Po_2017, Fidkowski_2019}, 
discrete quantum field theory~\cite{Brun_2020}, 
and the classification and characterization of symmetry protected topological (SPT) 
phases~\cite{Haah_2022, beyond, Shirley_2022}.
Many of these condensed matter applications 
hinge on the existence of \emph{nontrivial} QCA, 
that is, QCA which are not just low-depth circuits of local unitary gates.%
\footnote{%
Here low-depth means circuit depth that remains finite in the thermodynamic limit of infinite system size.}
The classification of such nontrivial QCA was obtained in one dimension by~\cite{GNVW},
where it was shown to be fully captured by an invariant equal to a rational number.
In the case of qudit degrees of freedom with prime number dimension,
this invariant reduces to an integer, corresponding simply to lattice translations.

In dimensions greater than one, the story is richer and less well understood.
A more useful notion of nontrivialness here 
is given by the so-called blending equivalence relation: 
a QCA $\alpha$ is nontrivial if one cannot consistently define a QCA that interpolates, 
or ``blends,'' between $\alpha$ in one region of space 
and the identity in a sufficiently distant complementary region.  
Blending is a coarser equivalence relation, 
with all translations parallel to blending surface
being trivial under blending in $d>1$.
It was shown in \cite{Freedman_2020} 
that all $2$-dimensional QCA are trivial under blending (see also \cite{haah2021clifford}).
However, in three dimensions various QCA have been constructed and argued to be nontrivial.
In particular, ref.~\cite{Haah_2022} constructed a QCA 
which disentangles the so-called three-fermion Walker--Wang Hamiltonian, 
which has the property that its surface state realizes the three-fermion topological order
entirely using local commuting projector terms in the Hamiltonian.
The three-fermion Walker--Wang Hamiltonian is an example of an SPT phase of time reversal 
that is beyond the cohomology classification, 
but appears in the cobordism classification of~\cite{kapustin2014symmetry},
with action $\frac{1}{2}\int w_2^2+w_1^4$, 
where $w_2$ is the second Stiefel--Whitney class of the spacetime tangent bundle.
Ref.~\cite{Haah_2022} used the chiral nature of the three-fermion topological order 
when realized strictly in two dimensions, 
and in particular the conjectured impossibility of realizing it 
with a strictly two-dimensional commuting projector Hamiltonian~\cite{Kitaev_2006},
to show that this QCA is nontrivial.
This argument was extended to other Walker--Wang models 
based on other chiral anyon theories in~\cite{haah2021clifford,Shirley_2022},
where a general classification of three dimensional QCA 
based on the Witt classification of such anyon theories was also proposed.

In this paper, we generalize along a different direction: 
namely, to QCA that disentangle higher dimensional generalizations 
of the $3$-fermion Walker--Wang model.  
Our motivation comes partly from Ref.~\cite{fidkowski2023pumping}, 
where a simple exact form was found for the unitary that realizes the $3$-fermion QCA.
Namely, it was found that the corresponding unitary $U^{3F WW}$ 
has a simple form when the underlying bosonic spin Hilbert space 
is viewed as the Hilbert space of a fermion coupled to an emergent $\Z_2$ gauge field%
\footnote{%
	This is the three dimensional analogue of Kitaev's honeycomb model construction~\cite{Kitaev_2006}; 
	the lattice level bosonization was based on \cite{chen2019bosonization}, as explained below.
}: 
given two stacked copies of such a Hilbert space, 
referred to as copy $1$ and copy $2$, the unitary $U^{3F WW}$ has eigenvalues $\pm 1$, 
with the $-1$ eigenspace spanned by states where the $\Z_2$ gauge fluxes of copy $1$ and copy $2$ 
have odd mutual linking number.
Such a mod $2$ linking number may be written in the Chern--Simons form $\int a \,\rd b$, 
where $a$ and $b$ are the $\Z_2$ gauge fields of the two copies.  
It was also shown in \cite{fidkowski2023pumping} 
that given even just a single copy, 
the unitary $(-1)^{\int a \, \rd a}$ is also in the same universality class 
of the $3$-fermion QCA, 
{\it i.e.}, it is equivalent to the other unitary up to a shallow circuit.

Motivated by the suggestive Chern--Simons form of these unitaries,
in this work we propose a general correspondence,
in arbitrary dimensions, 
between bosonic SPT phases of time reversal, as classified by unoriented cobordism, and QCA.
To do this, we first generalize the decomposition of the bosonic spin Hilbert space 
as a fermion coupled to an emergent $\Z_2$ gauge field.
Following techniques of \cite{chen2019bosonization, ChenTata, ChenPRR}, 
for any $n<d$, and any sequence $\{i_s\}$, $s=1,\ldots,r$ with $\sum i_s= n+1$, $i_s \geq 2 i_{s+1}$ for $1\leq s\leq r-1$, and $i_r \leq d-n$ we write a certain bosonic spin Hilbert space in $d$ dimensions as the Hilbert space of an $n$-form $\Z_2$ gauge field 
coupled to (extended) $n-1$ dimensional objects 
which can be thought of, in a certain sense, as fermions.
We emphasize that this is an exact re-writing of a lattice spin model Hilbert space;
the various gauge fields live on appropriate dimensional plaquettes of the lattice.
We then stack $m$ different such Hilbert spaces, 
with $n \in \{n_1,\ldots,n_m\}$, 
such that $\sum_j (n_j +1) = d+1$.
Denoting the corresponding $n_j$-form $\Z_2$ gauge field as $a_j$, 
we then define the unitary
\begin{align*}
	U^{a_1\,\rd a_2 \ldots \rd a_m}_{\bf{i}} \equiv (-1)^{\int a_1 \cup \rd a_2 \ldots \cup \rd a_m}
\end{align*}
where ${\bf{i}}$ packages the information in the sequences $\{i^j_1,\ldots, i^j_{r_j}\}$.
Our conjecture, which we substantiate with field theory calculations, is that by composing various QCA of this form, we can produce the ground state of any time-reversal protected SPT in the cobordism classification of SPT phases.  That is, we can produce any SPT phase with action of the form $\int w_{t_1} \ldots w_{t_m}$ where the $w_n$ are Stiefel--Whitney classes.
By construction, these states are invariant under a time reversal symmetry 
that can be taken to be complex conjugation in a natural basis.
We refer to this conjecture as the Stiefel--Whitney QCA (SW-QCA) correspondence.

This conjecture ties together several known examples of QCA.
The three-fermion QCA corresponds to $1$-form gauge fields $a_1$ and $a_2$ and $r_1=r_2=1$, so the resulting action is $\int \left(w_2^2 + w_1^4\right)$ as expected.
If we instead take $a_1$ to be a $1$-form and $a_2$ to be a $2$-form, with $\{i_s\}$ sequences $\{2\}$ and $\{2,1\}$ respectively, we expect the resulting QCA to produce the absolutely stable invertible $4+1$-dimensional phase 
with action $\int w_2 w_3$~\cite{Fidkowski_gravitational, Chen_2023anomaly}. 
A QCA producing this phase was indeed constructed in~\cite{Chen_2023anomaly},
and we expect that it is equivalent to the one in our conjectured correspondence. 

There are also examples involving the first Stiefel--Whitney class $w_1$,
where an additional subtlety arises.
Here the $0$-form gauge field is to be interpreted as an Ising degree of freedom 
with a global $\Z_2$ symmetry, 
and the state produced by our QCA can be considered an SPT of this unitary $\Z_2$ symmetry 
together with time reversal.%
\footnote{%
	In general, the state produced by our QCA 
	will also be an SPT of all of the $n_j$-form symmetries, 
	but we do not exploit this directly in the current work.
}
For example, for $a$ a $0$-form the QCA $U^{a\,\rd a \ldots \rd a}_{\bf{i}}$ 
will produce the in-cohomology SPT of time reversal for odd $d$, 
with action $\int w_1^d$, 
whereas for even $d$ it will produce the in-cohomology SPT of unitary $\Z_2$ symmetry, 
with action $\int A^d$.

Another example is the QCA $U^{a_1\,\rd a_2 \,\rd a_3}_{\bf{i}}$ 
with $a_1$ a $0$-form $\Z_2$ gauge field ({\it i.e.}, an Ising spin degree of freedom) 
and $a_2$, $a_3$ 1-form $\Z_2$ gauge fields.  Here all three sequences $\{i_s\}$ are one-term sequences.  If we ignore time reversal and only consider the unitary $\Z_2$ symmetry, this will produce an SPT with action $\int A w_2^2$, 
which is the beyond cohomology $4+1$ dimensional phase of unitary $\Z_2$ symmetry 
constructed in~\cite{beyond}.
The QCA constructed here gives a simpler way to obtain this phase than the model in~\cite{beyond}.
We note that in this case the QCA is actually a circuit, 
and is nontrivial only as a unitary $\Z_2$-symmetry protected QCA.
In general, a QCA obtained from the SW-QCA correspondence 
is guaranteed to be nontrivial only as a symmetry protected QCA, 
when the corresponding SPT is nontrivial.
When such a QCA produces an absolutely stable invertible phase, 
such as the $w_2 w_3$ phase, 
it is guaranteed to be a nontrivial QCA without any symmetry requirements.
However, the reverse implication is not true in general: 
{\it e.g.}, the $3$-fermion QCA is nontrivial 
even though the phase it produces requires time reversal symmetry for its protection.  We conjecture that, in general, the QCA is non-trivial precisely when it creates an invertible phase which has the property that its partition function is not equal to $1$ on some orientable spacetime manifold.

The main set of new QCA examples constructed in detail in this work 
are $U^{a \,\rd b}_{\bf{i}}$ in an odd number of dimensions $d$, 
with $a$ and $b$ being $n=\frac{d-1}{2}$ forms and the sequences $\{i_s\}$ being one-term sequences.
These generalize the $3$-fermion QCA (which corresponds to $d=3$) to higher dimensions.
Furthermore, they are Clifford QCA, and we write them in a compact form 
in the polynomial formalism of \cite{haah2021clifford}.  
We also demonstrate that, in accordance with field theory expectations, 
these QCA are trivial in the case of $d = 1 \bmod 4$.
We also outline some strategies for showing that, 
for all odd $d$, the QCA are not Clifford-trivial.

In general, it is useful to compare the predictions 
made by the SW-QCA correspondence 
with the classification of qubit Clifford QCA~\cite{haah2024topological},
which is $\Z_2$ in all odd spatial dimensions (except for $1+1$d),
and $0$ for all other spatial dimensions.
Which of the $U^{a_1\,\rd a_2 \ldots \rd a_m}_{\bf{i}}$ QCA are Clifford?
We will see that implementing $U^{a_1\,\rd a_2 \ldots \rd a_m}_{\bf{i}}$ 
requires $(m-1)$-fold controlled Pauli gates, 
so in order for this QCA to be Clifford, we must have $m\leq 2$.
Furthermore, in the decomposition of a bosonic $d$-dimensional Hilbert space 
into that of an $n$-form gauge field coupled to ``fermionic'' $n-1$ dimensional objects 
that we use to define the QCA, 
the operators that measure the $n$-form flux can be Pauli operators only if $n < \frac{d}{2}$ and the sequence $\{i_s\}$ contains only one term.  Combining these two constraints yields $m=2$, $n_1=n_2=\frac{d-1}{2}$, with $d$ necessarily odd.  
We conjecture that these are precisely the nontrivial Clifford QCA predicted in~\cite{haah2024topological}.

We note that several of the results in this work are not rigorously proven and remain conjectures.  
For example, the field theory argument for the QCA-SPT correspondence involves starting with a lattice model wave-function and computing its partition function on non-trivial spacetime manifolds, which runs into the usual problems of relating the Hamiltonian and spacetime action formulations of invertible phases.  In the same vein, we do not rigorously define a lattice level quantized invariant 
for the statistics of the fermionic $n-1$ dimensional excitations in the sense of 
{\it e.g.}~\cite{Fidkowski_gravitational}.  Furthermore, we do not rigorously prove that the Clifford QCA 
that we construct are the nontrivial ones in the classification of~\cite{haah2024topological}.
We will be clear about what is proven and what is conjectured below, and hope that the conjectures lead to more work on this topic.   

The rest of this paper is organized as follows.  
In \cref{duality} we construct an exact lattice duality 
between a bosonic spin Hilbert space and an $n$-form gauge theory 
coupled to $n-1$ dimensional objects that can be interpreted as fermions.  
This duality generalizes the bosonization dualities of~\cite{chen2019bosonization, ChenTata, ChenPRR}.  
We find the hypercubic lattice formulation of higher cup products of \cite{ChenTata} 
to be especially useful here.  
In \cref{gen3FQCA} we use these dualities to construct some of our QCA and examine their properties.  
We focus mostly on the Clifford QCA $U^{a \,\rd b}_{\bf{i}}$ 
with $a$ and $b$ being $k$ forms in $d=2k+1$ spatial dimensions and ${\bf{i}}$ being one-term sequences.  
We use the polynomial formalism of~\cite{haah2021clifford} 
to explicitly write the matrices that represent these translation-invariant Clifford QCA.  
We prove that for even $k$ ({\it i.e.}, in spatial dimensions congruent to $1$ modulo $4$),
this QCA is trivial, albeit requiring non-Clifford gates to write it as a circuit.
We also discuss some methods and progress in relating this QCA 
to the nontrivial ones in the Clifford QCA classification of \cite{haah2024topological}.  
Finally, in \cref{SWQCA} we discuss aspects of the general SW-QCA correspondence.  In particular, we give field theory and lattice `derivations' of the correspondence, and discuss the ways in which they fall short of being an actual proof.

\section{Exact lattice duality between a bosonic spin Hilbert space and an $n$-form gauge theory coupled to $n-1$ dimensional `fermions'} \label{duality}

The $3d$ bosonization duality of \cite{chen2019bosonization} can be viewed as being encoded in the Walker--Wang model based on the pre-modular category $\{1,f\}$, where $f$ is a fermion \cite{ChenTata}.  Indeed, this Walker--Wang model is just a $3d$ fermionic $\Z_2$ gauge theory built on a bosonic spin Hilbert space.  In the bosonization interpretation, the plaquette terms of this model are viewed as constraints that forbid gauge flux (see eq. 72 of \cite{chen2019bosonization}), whereas the vertex terms measure the occupation number of the emergent fermions (eq. 71 of \cite{chen2019bosonization}).  The rest of the fermionic operator algebra is generated by short fermionic string operators, which hop the emergent fermions (eq. 71 of \cite{chen2019bosonization}).  See section VII D of \cite{ChenTata} for an explicit derivation.

Our goal in this section is to generalize this Walker--Wang model to a higher dimensional one describing $n-1$ dimensional objects in $d$ dimensional space.  Importantly, it is not just the ground state of this model that will be important to us, but rather the entire spectrum: we will use these eigenstates, specified by the locations of the $n-1$ dimensional excitations and the $n$-form gauge fluxes, as a convenient basis when defining the QCA.\footnote{This will be formalized in the fact that the terms of this generalized Walker--Wang model, together with some global holonomies, form a `separator' in the sense of
\cite{Haah_2022}.}

The $3d$ $\{1,f\}$ Walker--Wang model corresponds to $n=1,d=3$: its ground state is a superposition of closed $1$-dimensional electric flux loops in $3$ dimensions, with the amplitude of such an electric flux loop containing fermionic signs.  In the following section we first review basic facts about these fermionic signs, in a way that allows easy generalization to other $n$ and $d$.  Note that in the context of the Walker--Wang model, $d=3$ here refers to the spatial dimension.  However, since the amplitudes in the Walker--Wang model can be interpreted as braiding histories in $3$-dimensional spacetime, $d=3$ will be viewed as a {\emph{spacetime}} dimension in the fermionic sign discussion below.

\subsection{Fermions in $2+1$ dimensions} \label{3dferm}
  
Given a fermion loop world-line~$L$ in $2+1$ dimensions, 
its contribution to the path integral includes a sign determined by an extra piece of data,
namely the framing of~$L$.
This is a choice of orthonormal frame in the normal bundle~$\nu_L$ to~$L$.
Given an orientation of the ambient $2+1$-dimensional space-time,
which we assume to be~$\RR^3$,
and a choice of direction along~$L$,
$\nu_L$ becomes oriented, 
so a single non-vanishing section~$v$ of~$\nu_L$ determines a frame.
One particular way to obtain such a non-vanishing section of $\nu_L$ 
is to take a linear projection $\RR^3 \rightarrow \RR^2$ 
such that $L$ is immersed in $\RR^2$ under this projection.  
Viewing the $\RR^2$ as the `blackboard,'
the vector field `out of the board' 
gives rise to an everywhere non-vanishing section $v$ of $\nu_L$.

The standard way to compute the sign $s_{L,v}$ associated to $L$ and $v$ is to use the spin structure, which in the case of flat spacetime $\RR^3$ can be taken to be trivial.  On a general $d+1$ dimensional orientable spacetime manifold $M$, a spin structure is a consistent lift of the transition functions defining the frame bundle to a set of transition functions that define its double cover Spin bundle.  When we supplement $v$ with a unit vector along the curve $L$ to form a section of the frame bundle over $L$, there may be an obstruction to lifting this to a section of the Spin bundle over $L$; if so, $s_{L,v}=-1$, otherwise $s_{L,v}=1$.

However, for a convenient generalization of the notion of fermionic statistics to higher dimensional objects, we now describe an alternative way of computing $s_{L,v}$, following the method of \cite{ChenPRR}.  To determine the sign associated to~$L$ and~$v$, 
we view the $2+1$-dimensional spacetime $\RR^3$ as the boundary, at $x_0=0$, of a $3+1$-dimensional spacetime $\RR^4_{x_0 \geq 0}$, where $x_0$ is the coordinate perpendicular to the boundary, 
and write~$L$ as the boundary of a disk~$D$: $L=\partial D$.  
We take the interior of ~$D$ to be in the interior of the $\RR^4_{x_0 \geq 0}$, and locally perpendicular to the boundary $\RR^3$ near $L$, 
i.e. we require that its normal bundle $\nu_D$, when restricted to $L$, 
reproduces that of $L$: $\nu_D \big|_L = \nu_L$.  
Then
\begin{align*}
    s_{L,v} = (-1)^{\int_D w_2(\nu_D)}
\end{align*}
which measures the obstruction to extending the $1$-frame $v$ of $\nu_L$ to a $1$-frame of $\nu_D$.  
Here $w_2(\nu_D)$ is the second Stiefel--Whitney class of the bundle $\nu_D$. 

It is instructive to generalize the discussion from four dimensional flat spacetime 
to an arbitrary smooth four dimensional manifold $M$, without boundary for the moment.  
Let $f: S \rightarrow M$ be an embedding of a $2$-dimensional surface in~$M$, 
and view $S$ as the Poincare dual representative of a cohomology class $e \in H^2(M,\Z_2)$.  
We then have:
\begin{align*}
    \Sq^2(e) = f_*(w_2(\nu_S))
\end{align*}
where $\Sq$ is the Steenrod square operation.
Thus
\begin{align} \label{theabove}
    \int_M e \cup e = \int_M \Sq^2(e) = \int_M f_*(w_2(\nu_S)) = \int_S w_2(\nu_S)
\end{align}
If we now go back to the case of four dimensional flat spacetime with an $\RR^3$ boundary, 
and $D$ and $L$ as before, 
we can similarly introduce a $2$-form $e$ that represents $D$ 
in a Poincare dual fashion in $\RR^4$.  
Because $D$ is perpendicular to $\RR^3$ near~$L$, 
$e$ restricts to a $2$-form on $\RR^3$, 
and $e \big|_{\RR^3}$ is Poincare dual to $L=\partial D$.
Also $\rd e=0$, 
so $e= \rd \epsilon$ for a $1$-form $\epsilon$, where the differential is taken relative to the boundary.
Then
\begin{align*}
    e \cup e = \rd \epsilon \cup \rd \epsilon = \rd (\epsilon \cup \rd\epsilon)
\end{align*}
so that $e \cup e$ integrates out to the boundary, and we obtain, using \cref{theabove}:
\begin{align} \label{3dsign}
    s_{L,v} = (-1)^{\int_D w_2(\nu_D)} = (-1)^{\int_{\RR^3} \epsilon \cup \rd\epsilon}
\end{align}
There is a subtlety in the above derivation, 
namely that the choice of framing of $L$ necessary to define $\int_D w_2(\nu_D)$ 
is encoded in the definition of the lattice cup product $\int_{\RR^3} \epsilon \cup \rd\epsilon$. 
\cite{ThorngrenThesis, ChenTata}.  
Indeed, $\int_{\RR^3} \epsilon \cup d\epsilon$ is just the mod $2$ linking number of~$L$ 
with itself, which requires $L$ to be infinitesimally `pushed off' along some direction; 
this choice of direction is precisely the framing information.  We will later interpret the quantity in the exponent on the right hand side of \cref{3dsign} as the logarithm of the ground state wavefunction of a generalized Walker-Wang model, denoted $\psi_{2,3}(\epsilon)$, on an arbitrary branched simplicial spatial manifold $M_3$:
\begin{align} \label{3dsignp}
\psi_{2,3}(\epsilon) = \int_{M_3} \epsilon \cup \rd\epsilon
\end{align}
Here the subscript $2$ refers to the degree of the form $\rd \epsilon$, and $3$ refers to the spatial dimension.

\subsection{Generalization to $k$ dimensional objects in $d=2k+1$ spacetime dimensions} \label{2k1ferm}

We now generalize the discussion of the previous section 
to obtain a formula for the fermionic sign associated to a $k$-dimensional submanifold $L$ of $\RR^{2k+1}$,
viewed as a $k-1$ dimensional extended object moving through $2k+1$ dimensional spacetime 
(the previous section was the case $k=1$).  Thus in this section $n=k, d=2k+1$; we will save the slightly more involved case of general for \cref{ndferm} below.  Again, we emphasize that while $d$ will be the spatial dimension of the generalized Walker--Wang model, in this section it plays the role of a space-time dimension.  The generalized Walker--Wang model will be constructed in \cref{2k1WW} by starting with a superposition of all configurations in $d$ dimensions weighted by the fermionic sign derived here, and bootstrapping this ground state into a Hamiltonian.
 
The discussion proceeds completely analogously to \cref{3dferm}.
We again extend spacetime to $2k+2$ dimensions 
and take a $k+1$ manifold $D$ with boundary $L$, 
orthogonal to $\RR^{2k+1}$ at $L$, 
so that $\nu_D \big|_L = \nu_L$.  
We again assume that we have a single section (i.e. a $1$-frame) $v$ of $\nu_L$, 
obtained from e.g. a blackboard framing.  
Viewing this as a section of $\nu_D \big|_L$, 
we see that it is $1= (k+1) - (k+1-1) = \text{dim}\, \nu_D - (k+1-1)$ frame of $\nu_D \big|_{\partial D}$,
so the obstruction to extending it to a $1$-frame of $\nu_D$ 
is given by the $k+1$st Stiefel--Whitney class $w_{k+1}(\nu_D)$.  

Going through the same steps as before, 
we can introduce a $k+1$-form $e$ describing a $k+1$ dimensional surface 
embedded via a map $f$ in $2k+2$ dimensions, 
and obtain $e\cup e = {\text{Sq}}^{k+1}\left( e\right)=f_*(w_{k+1}(\nu_S))$.  
Specializing to flat $2k+2$ dimensional spacetime with $2k+1$ dimensional boundary 
and integrating out to this boundary, we see that the fermionic sign of $L$ and $v$, which we once again denote $s_{L,v}$, is
\begin{align} \label{sign2k1}
    s_{L,v} = (-1)^{\int_{\RR^{2k+1}} \epsilon \cup \rd \epsilon}
\end{align}
with $\rd \epsilon$ dual to $L$.  Again, there is an implicit choice of pushoff direction hidden in the cup product.  Again we will later interpret the quantity in the exponent on the right hand side of \cref{sign2k1} as the logarithm of the ground state wavefunction of a generalized Walker-Wang model, denoted $\psi_{k+1,2k+1}(\epsilon)$, on an arbitrary branched simplicial spatial manifold $M_{2k+1}$:
\begin{align} \label{3dsignp}
\psi_{k+1,2k+1}(\epsilon) = \int_{M_{2k+1}} \epsilon \cup \rd\epsilon
\end{align}
Here the subscript $k+1$ refers to the degree of the form $\rd \epsilon$, and $2k+1$ refers to the spatial dimension.


\subsection{Generalization to $n$ dimensional objects in $d$ spacetime dimensions,} \label{ndferm}

We now compute the fermionic sign for the general case of $n$-dimensional surfaces in $d$ flat spacetime dimensions.  Once again this idea is based on generalizing the method of \cite{ChenPRR}.  We will first construct a special case of this action, in the cases when $n<\frac{d}{2}$.  Extend spacetime to $d+1$ dimensions with a $d$ dimensional boundary, 
and extend $L$ to an $n+1$ dimensional object $D$ with $L=\partial D$.  
We have the dimension of $\nu_D$ equal to the dimension of $\nu_L$, equal to $d-n$.  
In order for $w_{n+1} (\nu_D)$ to be the obstruction that heralds a negative fermionic sign, we need a $d-2n$ frame of $L$, which we denote by ${\bf{v}}$.  
This can be obtained from choosing $d-2n$ directions 
instead of just one blackboard direction, 
and reflects the fact that higher cup products 
will appear in the amplitudes, as we shall see below.

Going through the same steps as in \cref{3dferm} and \cref{2k1ferm},
we introduce a $d-n$ form $e$ dual to an $n+1$ surface $S$ in $d+1$ dimensions, 
and obtain $ e\cup_{d-2n-1}  e={\text{Sq}}^{n+1}\left( e\right)=f_*(w_{n+1}(\nu_S))$.  
Again we write $e=\rd \epsilon$, 
and try to integrate out to the boundary.  Note that
\begin{align*}
    \rd \epsilon \cup_{d-2n-1} \rd \epsilon 
    &= 
    \rd(\epsilon \cup_{d-2n-1} \rd\epsilon) 
    + \epsilon \cup_{d-2n-2} \rd\epsilon + \rd \epsilon \cup_{d-2n-2} \epsilon \\
&= \rd(\epsilon \cup_{d-2n-1} \rd\epsilon) + \rd(\epsilon \cup_{d-2n-2} \epsilon)
\end{align*}
Thus we obtain the fermionic sign 
\begin{align} \label{signgen}
    s_{L,{\bf{v}}} = (-1)^{\int \left(\epsilon \cup_{d-2n-1} \rd\epsilon + \epsilon \cup_{d-2n-2} \epsilon\right)}
\end{align}
This reduces to the formula \cref{sign2k1} when $d=2n+1$ because in that case the $\epsilon \cup_{d-2n-2} \epsilon$ term in \cref{signgen} is not present.  Also, in the special case of $n=0$ and general $d$, $\epsilon$ is a $d-1$ form and $d\epsilon$ is a $d$ form, consistent with the fact that it is dual to $0$ dimensional objects.  Once again, we define for general branched simplicial manifolds $M_d$:
\begin{align} \label{signgenp}
\psi_{n,d}(\epsilon) = \int_{M_d} \left(\epsilon \cup_{d-2n-1} \rd\epsilon + \epsilon \cup_{d-2n-2} \epsilon\right).
\end{align}

The formula $ e\cup_{d-2n-1}  e={\text{Sq}}^{n+1}\left( e\right)$ suggests a natural generalization, valid for any $n<d$.  Namely, for any positive integer $r$, we take a sequence $\{i_s\}$, $s=1,\ldots,r$ with $\sum i_s= n+1$, $i_s \geq 2 i_{s+1}$ for $1\leq s\leq r-1$, and $i_r \leq d-n$, and consider the action
\begin{align} \label{Sqcomp}
S_{\{i_s\},d}(e)= \int_{M_{d+1}} {\text{Sq}}^{i_1} \circ \ldots \circ {\text{Sq}}^{i_r} \left(e\right)
\end{align}
The restriction $i_s \geq 2 i_{s+1}$ for $1\leq s\leq r-1$ amounts to the fact that the compositions of Steenrod squares of the above type form the so-called Cartan basis of mod $2$ cohomology operations.  The restriction $i_r \leq d-n$ is necessary so that the degree of each Steenrod square is not greater than that of the form it is acting on.  The action in \cref{Sqcomp} can be written out in terms of higher cup products and integrated out to the boundary upon setting $e = \rd \epsilon$, but we will not carry out this procedure in detail here.  We will call the result $(-1)^{\psi^{\{i_s\}}_{n,d}(\epsilon)}$.

There are two particular instances of this action that will prove useful in the subsequent discussion.  One is the case $r=1$, which we will denote by $S^{\text{Wu}}_{n,d}$:
\begin{align} \label{S:Wu}
S^{\text{Wu}}_{n,d}(e) = \int_{M_{d+1}} {\text{Sq}}^{n+1} \left(e\right) = \int_{M_{d+1}} v_{n+1} \cup e
\end{align}
Here $v_{n+1}$ is the degree $n+1$ Wu class.  This action is non-zero only for $n < \frac{d}{2}$.  The other case is when all of the $i_r$ are powers of $2$, i.e. they form the non-zero digits of the binary decomposition of $n+1$.  In that case we refer to $S_{\{i_s\},d}(e)$ as $S^{\text{SW}}_{n,d} (e)$, where SW stands for Stiefel-Whitney.  The reason for this is that, as we will see in \cref{iteratedsquares},
\begin{align}
S^{\text{SW}}_{n,d}(e) = \int_{M_{d+1}}\left(w_{n+1} + \ldots \right)\cup e
\end{align}
where $w_{n+1}$ is the degree $n+1$ Stiefel-Whitney class, and the dots denote products of strictly lower degree Stiefel-Whitney classes.  This action is well defined for any $n<d$.

\subsection{$2k+1$ dimensional generalization of $3$ dimensional $\{1,f\}$ 
Walker--Wang model with $k$-form gauge field} \label{2k1WW}

We will now generalize the $\{1,f\}$ Walker--Wang model to a fermionic $k$-form $\Z_2$ gauge theory in $2k+1$ dimensions.  Recall that the ground state wave-function of the 3d $\{1,f\}$ Walker--Wang model is supported on $\Z_2$ loop configurations, which are viewed as `electric flux lines' in the gauge theory language.  The ground state amplitude of each such $\Z_2$ loop configuration is given by interpreting it as a worldline of a fermion in $2+1$ dimensional spacetime and computing the corresponding fermionic sign.  According to \cref{3dferm}, this amplitude is just $(-1)^{\int \epsilon \cup \rd\epsilon}$, where $\rd\epsilon$ gives the location of fermion worldlines.  In that derivation we could have worked with a simplicial decomposition of spacetime, but for our ultimate purpose of defining a QCA it will be much more convenient to work on a hyper-cubic lattice.  Fortunately, a convenient formulation of cup products and simplicial cohomology on hyper-cubic lattices was developed in \cite{ChenTata}, and we will use this formalism going forward.

Specifically, $\rd\epsilon$ is supported on $2$-plaquettes, which are dual to $1$-plaquettes supporting the fermion worldlines.  Of course in the full Hilbert space of the Walker--Wang model the fermion worldlines do not have to be closed, and are instead described by qubits on the such $2$-plaquettes $\square_2$, with operator algebra generated by the usual Pauli matrices $Z_{\square_2}, X_{\square_2}$.  $Z_{\square_2}=-1$ then means that the $1$-plaquette dual to $\square_2$ is occupied by a fermion worldline.  The utility of the gauge field $\epsilon$ is just in encoding the ground state amplitudes.

The natural generalization to $2k+1$ dimensions 
is that the ground state wave-function be supported on $k$-cycles in a hyper-cubic lattice, and that the amplitude should be $(-1)^{\int \epsilon \cup \rd\epsilon}$ (c.f. \cref{sign2k1}) 
where $\epsilon$ is now a $k$-form.  
$\rd\epsilon$ again gives the location of the electric surfaces in a dual fashion.  Specifically, we take qubit degrees of freedom living on $k+1$-plaquettes $\square_{k+1}$ of a $2k+1$ dimensional hypercubic lattice, and let $X_{\square_{k+1}}$ and $Z_{\square_{k+1}}$ be the Pauli operators acting on those qubits.  We interpret $Z_{\square_{k+1}}=-1$ as the $k$-cell dual to $\square_{k+1}$ being occupied by an electric surface.  The term that ensures that the electric surface is closed is $W_{\square_{k+2}}=1$, where
\begin{align*}
    W_{\square_{k+2}} &= \prod_{\square_{k+1}\subset \square_{k+2}} Z_{\square_{k+1}}
\end{align*}
To obtain the form of the plaquette term, we note that the plaquette term associated to a $k$-plaquette $\square_k$ simply adds a small loop of electric flux around the boundary of the $k+1$-plaquette dual to $\square_k$.  This just means that it shifts $\epsilon \rightarrow \epsilon + \square_k$, where here by abuse of notation we are interpreting $\square_k$ as the indicator cocycle on $\square_k$.  We can compute the change in the ground state amplitude under this shift:
\begin{align*}
    \int \left(\epsilon+\square_k\right) \cup \rd\left(\epsilon+\square_k\right) 
    - \int \epsilon \cup \rd\epsilon 
    &= 
    \int \left(\square_k \cup \rd\epsilon + \epsilon \cup \rd\square_k + \square_k \cup \rd\square_k\right) \\
    &= 
    \int \left(\square_k\cup \rd\epsilon + \rd\epsilon \cup \square_k + \square_k \cup \rd\square_k\right) \\
    &= 
    \int \rd\square_k \cup_1 \rd\epsilon + \int \square_k \cup \rd\square_k
\end{align*}
In the second equality above we performed an integration by parts, 
and in the last equality we used the property (see \cite{ChenTata} for a derivation in the hypercubic lattice setting):
\begin{align*}
    \rd(a \cup_1 b)= \rd a \cup_1 b + a \cup_1 \rd b + a \cup b + b \cup a
\end{align*}
Thus, the form of the plaquette operator that leaves the ground state wave-function invariant is $(-1)^{\int \square_k \cup \rd\square_k} G_{\square_k}$, where
\begin{align} \label{2k1G}
    G_{\square_k} 
    &= 
    \prod_{\square_{k+1} \supset \square_k} X_{\square_{k+1}} \left( \prod_{\square'_{k+1}} Z_{\square'_{k+1}} ^ {\int d \square_k \cup_1 \square'_{k+1}} \right)
\end{align}
The Walker--Wang Hamiltonian is then
\begin{align} \label{HWW}
    H^{WW}_{2k+1} = -\sum_{\square_{k+2}} W_{\square_{k+2}} - \sum_{\square_k} (-1)^{\int \square_k \cup \rd\square_k} G_{\square_k}
\end{align}
This is a Hamiltonian that describes $k-1$ dimensional electrically charged objects 
coupled to a $k$-form $\Z_2$ gauge field.  
In the case $k=1$, $d=3$ these objects are fermions.  
Considering the field theory derivation of this model, 
we expect that in higher dimensions these objects also have fermionic statistics, 
but we do not attempt to give a lattice definition of such statistics here.  

As in the case of $3$ dimensions, 
we can also write down analogues of short string operators 
which move the fermions but commute with the operators that measure the gauge fluxes:
\begin{align} \label{2k1U}
    U_{\square_{k+1}} 
    = 
    X_{\square_{k+1}} \prod_{\square'_{k+1}} Z_{\square'_{k+1}}^{\int \square'_{k+1} \cup_1 \square_{k+1}}
\end{align}
One can verify explicitly that $U_{\square_{k+1}}$ commutes with all $G_{\square_k}$, 
so that $U_{\square_{k+1}}$ does not alter the gauge fluxes.  
$U_{\square_{k+1}}$ does anti-commute with $W_{\square_{k+2}}$ 
for all $\square_{k+2}\supset \square_{k+1}$, 
so it creates a short electric flux loop.

We note that these definitions of $W$,$G$, and $U$ 
are natural generalizations of eqs. 30-32 of \cite{ChenTata}, 
which describe the bosonization of point-like fermions.  
In the bosonization interpretation, $G_{\square_k}=1$ are viewed as constraints, 
since what is being described in that case is a fermionic Hilbert space with no gauge fluxes.  In our interpretation, the $G_{\square_k}$ are simply gauge flux measurement operators.  We also note that the Hamiltonian in \cref{HWW} 
is an unfrustrated commuting Pauli Hamiltonian, 
and that specifying the location of the electric charge excitations (violations of $W_{\square_{k+2}}$) and gauge flux excitations (violations of $G_{\square_k}$), 
as well as global holonomies of the gauge field around any globally nontrivial cycles 
determines a state uniquely.  
Indeed, for any such specification of excitations and holonomies, there exists precisely one state, i.e. the terms in $H^{WW}_{2k+1}$ 
together with operators that measure gauge field holonomies 
around nontrivial $k$-cycles form a separator \cite{Haah_2022}.


\subsection{$d$-dimensional generalization of $\{1,f\}$ Walker--Wang model with $n$-form gauge field}
\label{genWW}

We can generalize the Walker--Wang construction to the case derived from $n-1$ dimensional objects moving in $d$ dimensional spacetime, i.e. to the case where the electric surfaces have some general dimension $n$.  As in section \cref{ndferm}, we first consider a special case of $S^{\text{Wu}}_{n,d}$, which is valid only when $n<\frac{d}{2}$.  The resulting Walker--Wang model will be in $d$ dimensional space, and will feature qubits defined on dual plaquettes $\square_{d-n}$; $\epsilon$ is then a $(d-n-1)$-form, and again $\rd \epsilon(\square_{d-n})\neq 0$ means that the plaquette dual to $\square_{d-n}$ is occupied by an electric surface.  The vertex term is now
\begin{align*}
    W_{\square_{d-n+1}} &= \prod_{\square_{d-n}\subset \square_{d-n+1}} Z_{\square_{d-n}}
\end{align*}
To get the plaquette term, we have to examine the variation of (c.f. \cref{signgen})
\begin{align} \label{01sign}
    \int \left(\epsilon \cup_{d-2n-1} d\epsilon + \epsilon \cup_{d-2n-2} \epsilon\right)
\end{align}
under $\epsilon \rightarrow \epsilon + \square_{d-n-1}$.  The $\epsilon$-dependent portion of this variation is
\begin{align*}
\int &\left(\square_{d-n-1} \cup_{d-2n-1} \rd\epsilon + \epsilon \cup_{d-2n-1} d\square_{d-n-1} + \square_{d-n-1}\cup_{d-2n-2} \epsilon + \epsilon \cup_{d-2n-2} \square_{d-n-1}\right)\\
&=\int\left(\square_{d-n-1} \cup_{d-2n-1} \rd\epsilon+ \rd\epsilon \cup_{d-2n-1} \square_{d-n-1}\right)\\
&=\int\left(d\square_{d-n-1} \cup_{d-2n} \rd\epsilon\right)
\end{align*}
where in the first equality we used the integration by parts formula for higher cup products, 
which absorbed the last two terms.  The portion of the variation independent of $\epsilon$ is $\int \left(\square_{d-n-1} \cup_{d-2n-1} d\square_{d-n-1} + \square_{d-n-1} \cup_{d-2n-2} \square_{d-n-1}\right)$.  Thus, defining
\begin{align} \label{ndplaquette}
    G_{\square_{d-n-1}} 
    &= 
    \prod_{\square_{d-n} \supset \square_{d-n-1}} X_{\square_{d-n}} \left( \prod_{\square'_{d-n}} Z_{\square'_{d-n}} ^ {\int \rd \square_{d-n-1} \cup_{d-2n} \square'_{d-n}} \right)
\end{align}
we obtain the general Walker--Wang Hamiltonian
\begin{align} \label{HWWgen}
    H^{WW}_{n,d} = -\sum_{\square_{d-n+1}} W_{\square_{d-n+1}} - \sum_{\square_{d-n-1}} (-1)^{\int \left(\square_{d-n-1} \cup_{d-2n-1} \rd\square_{d-n-1} + \square_{d-n-1} \cup_{d-2n-2} \square_{d-n-1}\right)} G_{\square_{d-n-1}}
\end{align}
One can likewise write down short surface operators that commute with the plaquette terms and move the electric excitations, but we will not need these in the rest of the paper.  The Walker--Wang Hamiltonian $H^{WW}_{n,d}$ describes electrically charged $n-1$ spatial-dimensional excitations coupled to an $n$-form gauge field, for $n < \frac{d}{2}$.

\subsubsection*{Special case of $n=0$:} \label{nequals0}
It is useful to examine in more detail the special case of $n=0$ and $d \geq 1$.  
Here the `electric surfaces' are $0$ dimensional, so the would-be excitations corresponding to their boundaries do not exist.
Nevertheless, the corresponding Walker--Wang model still makes sense.  
The $0$ dimensional `electric surfaces' can be viewed as Ising spins, living on $d$ dimensional cells.  
The plaquette terms simply hop these, 
and live on $d-1$ dimensional cells.  
For $d=1$, these plaquette terms are of the form $Y_{j-1} Y_j$, where $j$ is an integer index labeling the lattice sites on a line.  This can be seen from the explicit formula in \cref{ndplaquette}, or by arguing as follows: the sign structure \cref{01sign} that the plaquette has to preserve is $\int \epsilon \cup \rd \epsilon$, which is just $-1$ raised to one half of the number of electric surfaces (which occur when $\rd \epsilon \neq 0$).  This sign flips precisely when a plaquette term creates or destroys two electric surfaces, as opposed to hopping an electric surface from one site to the other, i.e. when $Z_{j-1} Z_j = -1$.  Hence the plaquette term is $X_{j-1} X_j (-Z_{j-1} Z_j) = Y_{j-1} Y_j$.  

For $n=0$ with general $d$, we still have a $0$-form global $\Z_2$ symmetry, together with complex conjugation time reversal symmetry.  The generalized QCA that we construct based on such a model will then have both of these symmetries.  

\subsubsection*{Most general case, arbitrary $n$, $d$} \label{mostgen}

Let us now consider the most general situation, corresponding to the $d+1$ dimensional bulk action ${\text{Sq}}^{i_1} \circ \ldots {\text{Sq}}^{i_r} \left( e\right)$ in \cref{Sqcomp}.  As mentioned above, this action can also be integrated out to the $d$ dimensional boundary upon setting $e = \rd \epsilon$, simply by expressing the Steenrod squares in terms of higher cup products, and repeatedly applying the identity
\begin{align*}
\rd\left(a\cup_j b\right) = \rd a \cup_j b + a \cup_j \rd b + a \cup_{j-1} b + b \cup_{j-1} a
\end{align*}
By varying the resulting wavefunction for $\epsilon$ with respect to local changes of $\epsilon$, we can write down plaquette terms analogous to the $G_{\square_{d-n-1}}$ above.  Together with the vertex terms $W_{\square_{d-n+1}}$, which again just ensure that the Poincare dual to $\epsilon$ has no boundary, one can write down a corresponding Walker-Wang type Hamiltonian.  This Hamiltonian results in a topological order which again has $n-1$ dimensional `charge' excitations and $d-n-1$ dimensional `gauge flux' excitations.  We will not write the Hamiltonian out in detail in this paper, since this most general case does not play a role in any of the specific examples that we study.  It does, however, play a role in the Stiefel-Whitney - QCA correspondence, in that we need these most general models in order to ensure that we can produce any cobordism SPT phase of time reversal by acting with appropriate QCAs on a product state.

Note that $\int_{M_{d+1}} {\text{Sq}}^{i_1} \circ \ldots {\text{Sq}}^{i_r} \left( e\right)$ is simply the bulk action for the Walker-Wang Hamiltonian that is being discussed here, with the special case of $r=1$ and $n<\frac{d}{2}$ corresponding to the Hamiltonian in \cref{HWWgen}.  It is worth examining this action in some special cases.  First, consider $r=1$, $n=1$, and general $d$.  Then $e$ is a $d-1$ form, and
\begin{align*}
  S^{\text{Wu}}_{1,d}(e) = S^{\text{SW}}_{1,d}(e)= \int_{M_{d+1}} {\text{Sq}}^2 \left( e\right) =   \int_{M_{d+1}} v_2 \cup  e =   \int_{M_{d+1}}(w_2 + w_1^2) \cup  e
\end{align*}
where $v_k$ are the Wu classes.  If we restrict to orientable manifolds $M_{d+1}$, the action is just $  \int_{M_{d+1}} w_2 \cup  e$, which is consistent with our understanding that in bosonic systems, emergent point-like fermions see the second Stiefel Whitney class, which is the obstruction to the existence of a spin structure, as a gauge flux.  Now consider $n=2$, and general $d\geq 4$.  This time $e$ is a $d-2$-form, and there are two possible actions.  One is 

\begin{align*}
  S^{\text{Wu}}_{2,d}(e) = \int_{M_{d+1}} {\text{Sq}}^3 \left( e\right) =   \int_{M_{d+1}} v_3 \cup  e =   \int_{M_{d+1}} w_1 w_2 \cup  e
\end{align*}
while the other is
\begin{align} \label{Sq2Sq1}
  S^{\text{SW}}_{2,d}(e) = \int_{M_{d+1}} {\text{Sq}}^2 \circ {\text{Sq}}^1 \left( e\right) =   \int_{M_{d+1}} (w_3 + w_1^2) \cup  e
\end{align}
On orientable manifolds, this second action is just $  \int w_3 \cup  e$, which is a natural generalization of the notion of fermions to one-dimensional loop-like excitations \cite{Thorngren_2015}.  The equality in \cref{Sq2Sq1} is derived in \cref{iteratedsquares}.

\section{$2k+1$ dimensional generalization of the $3$-fermion QCA}
\label{gen3FQCA}

We now again turn to the special case $n=k$, $d=2k+1$ discussed in detail in \cref{2k1WW}, and construct the corresponding QCA.
To define our QCA, we will start with $2$ copies of the generalized $\{1,f\}$ WW model \cref{HWW}, which we assume live on two interpenetrating (i.e. dual) hyper-cubic lattices in $2k+1$ spatial dimensions.  Specifically, the degrees of freedom are qubits on $(k+1)$-plaquettes in both of these lattices.  To simplify notation, we will pick one of these two lattices, and assign qubits to both $(k+1)$-plaquettes and $k$-plaquettes in this one lattice.  The latter are precisely the dual $(k+1)$-plaquettes, and will be referred to as `dual' qubits forming the `dual' Hilbert space.  We will use the notation 
$X_{\square_{k+1}}, Z_{\square_{k+1}}, \tX_{\square_{k}}, \tZ_{\square_{k}}$ 
for the Pauli operators on these qubits; the tilde signifies the dual.
We recapitulate the definitions of $G_{\square_k}$ \cref{2k1G} and $U_{\square_{k+1}}$ \cref{2k1U} and write the dual versions:

\begin{align}
    G_{\square_k} &= \prod_{\square_{k+1} \supset \square_k} X_{\square_{k+1}} \left( \prod_{\square'_{k+1}} Z_{\square'_{k+1}} ^ {\int \rd \square_k \cup_1 \square'_{k+1}} \right)\\
    \tG_{\square_{k+1}} &= \prod_{\square_{k} \subset \square_{k+1}} \tX_{\square_{k}} \left( \prod_{\square'_{k}} \tZ_{\square'_{k}} ^ {\int \rd \hsquare_{k+1} \cup_1 \hsquare'_{k}} \right)\\
    U_{\square_{k+1}} &= X_{\square_{k+1}} \prod_{\square'_{k+1}} Z_{\square'_{k+1}}^{\int \square'_{k+1} \cup_1 \square_{k+1}}\\
    \tU_{\square_{k}} &= \tX_{\square_{k}} \prod_{\square'_{k}} \tZ_{\square'_{k+1}}^{\int \hsquare'_{k} \cup_1 \hsquare_{k}}
\end{align}
Here $\hsquare_j$ is the $d-j$-face Poincare dual to $\square_j$, 
with Poincare duality being defined by 
a shift in the $\left(\frac{1}{2},\ldots,\frac{1}{2}\right)$ direction 
and the usual intersection; 
see below for more details.  
Note that $[U,G]=[U,\tG]=[\tU,G]=[\tU,\tG]=0$ for all choices of subscript.

We will define our QCA $\alpha^{WW}_{2k+1}$ purely in terms of its action on local operators, 
and will explicitly verify that it is an algebra homomorphism.  
Nevertheless, it is illuminating to first define the associated unitary operator $U^{WW}_{2k+1}$.  
For concreteness, assume that we are on a large $2k+1$ dimensional torus.  
Recall that we are interpreting both the original Hilbert space and its dual 
as the Hilbert spaces of $k$-form gauge fields coupled to $k-1$ dimensional extended objects.  
Let us denote these two gauge fields as $a$ and $b$; 
then ${\hat{da}}(\square_{k}) = 1-2G_{\square_k}$ 
and $db(\square_{k+1})=1-2\tG_{\square_{k+1}}$.  
For consistency, $a$ lives on the $k$-faces of the dual lattice, 
so that $da$ lives on the $k+1$ faces of the dual lattice, 
which are $k$-faces of the original lattice.  
The associated unitary is then defined as
\begin{align} \label{UWW}
    U^{WW}_{2k+1} = (-1)^{\int a \cup \rd b}
\end{align}
Let us unpack this definition a little.  
First, note that $a$ is not well defined on the lattice; only $\rd a$ is well defined.  
Thus, a priori, it is not clear that this is a good definition.  
However, we can view $\int a \cup \rd b$ as 
computing the mod $2$ linking number of the gauge fluxes $\rd a$ and $\rd b$.
This mod $2$ linking number is well defined.
To see that there is no ambiguity, 
we just need to show that the expression $\int a \cup \rd b$ does not change, 
modulo $2$, 
when we shift $a$ by a closed $k$-form.  
Equivalently, we need to show that $\int a \cup \rd b=0$ 
when $a$ is closed, and this just follows from integration by parts.

Second, we note that, strictly speaking, 
the right hand side of \cref{UWW} 
should be written in terms of $G_{\square_k}$ and $\tG_{\square_{k+1}}$ 
rather than $a$ and $b$.  
We should view \cref{UWW} as saying that $U^{WW}_{2k+1}$ 
is diagonalized by the eigenstates of the generalized $\{1,f\}$ Walker--Wang models, 
and the eigenvalue is just the mod $2$ linking number of the fluxes.

Let us now work out the action of $U^{WW}_{2k+1}$ on local operators:
\begin{align*}
    \alpha^{WW}_{2k+1} \left({\cal{O}}\right) = \left(U^{WW}_{2k+1}\right)^\dagger \,{\cal{O}}\, U^{WW}_{2k+1}
\end{align*}
First, take $\tZ_{\square_k}$.  Note that it anti-commutes with $\tG_{\square_{k+1}}$ if and only if $\square_k \subset \square_{k+1}$.  Thus, it creates a short loop ${\tilde{l}}$ of dual magnetic flux (that is, flux of $db$) through all such $\square_{k+1}$.  Then the change in the linking number of the flux and dual flux is the linking number of $a$ and ${\tilde{l}}$.  This linking number is non-zero if and only if $\square_k$ is occupied by a magnetic flux, i.e. $G_{\square_k}=-1$.  Thus:
\begin{align} \label{alphatZ}
    \alpha \left(\tZ_{\square_k}\right) = \tZ_{\square_k} G_{\square_k}
\end{align}
Similarly,
\begin{align} \label{alphaZ}
    \alpha \left(Z_{\square_{k+1}}\right) = Z_{\square_{k+1}} \tG_{\square_{k+1}}
\end{align}
To get the action of $\alpha^{WW}_{2k+1}$ on the Pauli $X$ operators, we use the fact that the $U$ and $\tU$ operators commute with $U^{WW}_{2k+1}$, simply because they commute with all of the $G$ and $\tG$ operators.  Now, using
\begin{align*}
    X_{\square_{k+1}} = U_{\square_{k+1}} \left(\prod_{\square'_{k+1}} Z_{\square'_{k+1}}^{\int \square'_{k+1} \cup_1 \square_{k+1}}\right)
\end{align*}
we see that
\begin{align} \label{alphaX}
    \alpha^{WW}_{2k+1} \left(X_{\square_{k+1}}\right) &= U_{\square_{k+1}} \left(\prod_{\square'_{k+1}} \alpha^{WW}_{2k+1}\left(Z_{\square'_{k+1}}\right)^{\int \square'_{k+1} \cup_1 \square_{k+1}}\right) \\
    &= 
    X_{\square_{k+1}} \prod_{\square'_{k+1}} \left(Z_{\square'_{k+1}} \alpha^{WW}_{2k+1}\left(Z_{\square'_{k+1}}\right)\right)^{\int \square'_{k+1} \cup_1 \square_{k+1}}\\
    &= 
    X_{\square_{k+1}} \prod_{\square'_{k+1}} \tG_{\square'_{k+1}}^{\int \square'_{k+1} \cup_1 \square_{k+1}}.
\end{align}
Similarly,
\begin{align} \label{alphatX}
    \alpha^{WW}_{2k+1} \left(\tX_{\square_{k}}\right) &= \tX_{\square_k} \prod_{\square'_k} G_{\square'_k}^{\int \hsquare'_k \cup_1 \hsquare_k}
\end{align}
This uniquely defines the action of the QCA on operators.  In the next subsection, we write out this action in the polynomial formalism.

\subsection{Writing the $2k+1$ dimensional Walker Wang QCA in the polynomial formalism} \label{polyQCA}

\cref{alphaZ}, \cref{alphatZ}, \cref{alphaX}, and \cref{alphatX} encode all of the information about our QCA.  However, since this is a Clifford QCA, it is also useful to explicitly write down the matrix representation of the QCA in the polynomial formalism, which is a compact way of encoding translationally invariant Clifford operators - see section IV A of \cite{Haah_2022} for a review.  To do this, we define `original' $k$ and $k+1$ cells to be ones which have all coordinates either $0$ or ranging in $[0,1]$.  All other cells are translates of these, and each cell has a unique original translate.  Let $R = {2k+1 \choose k} = {2k+1 \choose k+1}$.  The QCA $\alpha^{WW}_{2k+1}$ will then be a $4R$ by $4R$ matrix valued in the polynomial ring in indeterminates $x_1,\ldots, x_{2k+1}$ over the field with $2$ elements.  Before writing down this matrix, let us first define some smaller, $R$ by $R$ matrices that encode standard co-chain operations.  All of these are co-chain operations in the hyper-cubic lattice cohomology formalism of \cite{ChenTata}.

{\bf{Boundary operation:}} Take $\square_{k+1}$ and $\square_{k}$ to be original.  We will use the notation $\square_{k+1}=[j_1<\ldots<j_{k+1}]$ to mean that $\square_{k+1}$ is extended along the directions $j_1$ through $j_{k+1}$.  Then the boundary operation is encoded in the matrix $B_{\square_k,\square_{k+1}}$, which is non-zero precisely when $\square_k$ is of the form $[j_1<\ldots<{\hat{j_\mu}}<\ldots<j_{k+1}]$, and is equal to $(1+x_{j_\mu})$ in this case \cite{ChenTata}.  Here the hat means that $j_\mu$ is omitted.

{\bf{Coboundary operation:}} The coboundary is encoded in a matrix $\Delta_{\square_{k+1},\square_k}$, defined by $\Delta_{\square_{k+1},\square_k} = \left(B_{\square_k, \square_{k+1}}\right)^*$, i.e. $\Delta = B^\dagger$.  Here the $*$ operation takes $x_j \rightarrow x_j^{-1}$ for all $j$.

{\bf{$\cup_1$ product:}} Next we encode the $\cup_1$ product as an $R$ by $R$ matrix $W$.  $W_{\square'_{k+1}, \square_{k+1}}$ is a monomial-valued entry which is $0$ if $\int T_{\vec{\alpha}} \square'_{k+1} \cup_1 \square_{k+1} = 0$ for all $\vec{\alpha}$ (here $T$ is the discrete translation operator on the lattice), and otherwise it is $x^{\vec{\alpha}} \equiv x_1^{\alpha_1} \ldots x_{2k+1}^{\alpha_{2k+1}}$, where $\vec{\alpha}$ is the unique integer valued vector such that $\int T_{\vec{\alpha}} \square'_{k+1} \cup_1 \square_{k+1} \neq 0$.  Note that in this case, the $2k+1$ chain $T_{\vec{\alpha}} \square'_{k+1} \cup_1 \square_{k+1}$ is non-zero on precisely one $2k+1$ cell, namely the original one).

Explicitly, using the definition in \cite{ChenTata}, we note that two cells $\square_{k+1}' = [j_1'<\ldots<j_{k+1}']$ and $\square_{k+1}=[j_1<\ldots<j_{k+1}]$ will have $W_{\square'_{k+1}, \square_{k+1}} \neq 0$ if and only if there exists a unique $m \in \{1,\ldots,2k+1\}$ such that $m=j_\mu=j'_{\mu'}$ for some $\mu,\mu'$.  In other words, $\{j_1',\ldots,j_{k+1}'\} \cap \{j_1,\ldots,j_{k+1}\} = \{m\}$.  In this case, letting ${\hat{i}}$ be the unit $2k+1$ dimensional vector in the $i$'th direction ($i=1,\ldots,2k+1$), we have
\begin{align*}
{\vec{\alpha}} = \sum_{j'_{\mu'} < m} {\hat{j'_{\mu'}}} - \sum_{j_\mu>m} {\hat{j_\mu}}
\end{align*}

{\bf{Duality:}} $D$ will be a matrix that encodes duality, which is just Poincare duality followed by a shift by $-\left(\frac{1}{2},\ldots, \frac{1}{2}\right)$.  $D_{\square_k,\square_{k+1}}$ will be non-zero if and only if $\square_k$ is the unique original cell that has a translate that is dual to the original cell $\square_{k+1}$.  In that case, we define $D_{\square_k,\square_{k+1}} = x^{\vec{\alpha}}$, where $\vec{\alpha}$ is the unique vector such that $T_{\vec{\alpha}} \square_k$ is dual to $\square_{k+1}$.  Explicitly, for $\square_{k+1} = [j_1 < \ldots < j_{k+1}]$, $\square_k = [i_1 < \ldots < i_k]$, where $\{i_1,\ldots,i_k\}$ is the complement of $\{j_1,\ldots,j_{k+1}\}$ in $\{1,\ldots,2k+1\}$, and $\vec{\alpha} = - \left({\hat{i_1}} + \ldots + {\hat{i_k}}\right)$.

{\bf{Dual $\cup_1$ product:}} We also want to encode a version of the $\cup_1$ product defined on dual $k$-cells, namely $\int \hsquare'_k \cup_1 \hsquare_k$.  To do this, we do Poincare duality and a shift by $-\left(\frac{1}{2},\ldots, \frac{1}{2}\right)$ to go from $k$-cells to $k+1$ cells, act with $W$, and then undo this.  More precisely, we define
\begin{align*}
{\tilde{W}} = D W D^\dagger
\end{align*}

{\bf{Matrix form of the QCA:}} Using the above definitions, we now write down $M$, the matrix form of our QCA.  This just follows directly from  \cref{alphaZ}, \cref{alphatZ},\cref{alphaX},\cref{alphatX} and the above matrix representations of the cohomology operations found in those formulas:
\begin{align}
\begin{pmatrix} \label{QCAmatrix}
{\vmathbb{1}} & W^\dagger B^\dagger & 0 & W^\dagger B^\dagger {\tilde{W}} \\
{\tilde{W}}^\dagger B & {\vmathbb{1}} & {\tilde{W}}^\dagger B W & 0 \\
0 & B^\dagger & {\vmathbb{1}} & B^\dagger {\tilde{W}} \\
B & 0 & BW & {\vmathbb{1}}
\end{pmatrix}
\end{align}
If we define the matrix $\lambda$ encoding the commutation relations
\begin{align}
\lambda = \begin{pmatrix}
0&0&{\vmathbb{1}}&0\\
0&0&0&{\vmathbb{1}}\\
{\vmathbb{1}}&0&0&0\\
0&{\vmathbb{1}}&0&0
\end{pmatrix}
\label{eq:lambda}
\end{align}
we can explicitly check that $M^\dagger \lambda M = \lambda$, so $M$ defines a QCA.  The proof uses $B(W+W^\dagger)B^\dagger=0$ and $B^\dagger\left({\tilde{W}} + {\tilde{W}}^\dagger\right)B=0$, which follow from the fact that $\int\left(da \cup_1 db + db \cup_1 da\right)=0$.  We also explicitly checked this numerically for $k=1,2$, i.e. $d=3,5$.

\subsection{Algorithms for descent maps}

For Clifford QCA, there is a classification theorem~\cite{haah2024topological},
which shows that with qubits $\CC^2$ 
there are exactly two classes (one trivial and the other not)
in every odd spatial dimension $\ge 3$ modulo Clifford circuits and shifts.
This classification is achieved by a sequence of dimensional descent maps
that relate each class of Clifford QCA 
to a Witt equivalence class of certain bilinear or quadratic forms.
The proof is constructive in that the descent maps are explicit,
and the $\ZZ_2$ invariant for each Clifford QCA 
is found by the final result of the descent map in zero dimension.
If $d$ is odd, then
we reach a nonsingular quadratic form in zero variables,
which is a usual quadratic form over $\FF_2$.
The Arf invariant of this quadratic form 
is the $\ZZ_2$ invariant of the original Clifford QCA.
It is somewhat subtle why we require $d \ge 3$,
but in short this is 
because the theory of quadratic forms and symmetric forms 
are not the same over a ring of characteristic~$2$,
and physically it is related to time reversal symmetry
whether a real operator is mapped to a real operator by the QCA.
This distinction will appear in the algorithm below.
See \cite{haah2024topological} for details.

We have implemented the descent maps
on a symbolic computer algebra system,
and calculated the $\ZZ_2$ invariant of our Clifford QCA in spatial dimension~$d=3$.
The invariant is nonzero.
The calculation for $d=5$, though finite, 
took a prohibitively long time that we could not report the result here;
we were able to execute the first descent map fully,
but we needed four more.
This of course does not indicate any fundamental obstacle.
We note only that a naive algorithm's computational time complexity 
to determine the $\ZZ_2$ invariant of a Clifford QCA in $d$ dimensions,
is at least doubly exponential in~$d$ 
because Gr\"obner basis algorithms~\cite{Dube1990} are used as a subroutine.%
\footnote{%
	In some cases, Gr\"obner basis finding is not too slow~\cite{Kera2023groebner}.
}

Below we give an algorithm for the dimensional descent maps,
deferring all the reasons for any assertion in the algorithm
to~\cite{haah2021clifford,haah2024topological}.
We use one blackbox subroutine that outputs a basis of a free module
over a Laurent polynomial ring~$R_d = \FF_2[x_1^\pm,\ldots,x_d^\pm]$
given a matrix whose columns generate the module.
(In our context, a free basis of a module 
is a collection of column ``vectors''
that can express any generator using $R_d$-linear combinations
and the only way to express zero using basis elements is the zero combination.)
This is generally possible by syzygy calculations using Gr\"obner bases
for Laurent polynomial rings~\cite{Park1995}.
However, in our computer algebra calculation, 
we did not use this general algorithm 
because it appeared to take too long.
Instead we used an {\it ad hoc} calculation 
that reduced the number of generators for a module.
After all, the rank of a free module can be calculated 
by minors (determinants of submatrices) of a generator matrix without knowing a free basis,
and once we find a generating set of size equal to the rank,
we know it has to be free for situations of interest.

\paragraph{From QCA to hermitian forms.}

Any $d$-dimensional translation-invariant Clifford QCA over qubits ($\CC^2$)
is encoded in a $2q \times 2q$ matrix~$Q$ such that $Q^\dag \lambda Q = \lambda$
over the Laurent polynomial ring $R_d = \FF_2[x_1^\pm,\ldots,x_d^\pm]$,
where $q$ is the number of qubits per lattice point,
$\dag$ is a matrix operation that 
takes the transpose and the entry-wise involution that inverts every variable,
and $\lambda$ is defined in \eqref{eq:lambda}.
Since the matrix $Q$ has only finitely many entries,
the exponents of $x_d$ across all entries 
({\it e.g.}, $-3$ in $x_1^5 x_2^7 x_d^{-3}$)
are contained in some integer interval $[a,b]$.
We transform $Q$ to another matrix 
such that the exponent of $x_d$ is either~$0$ or~$1$.
This is done by \emph{translation} and \emph{coarse-graining} as follows.
First, multiply $Q$ by a diagonal matrix $x_d^{-a} I_{2q}$.
(There is no difference whether the diagonal matrix is multiplied on the left or right.)
The resulting matrix~$Q'$ has all the $x_d$-exponents nonnegative
with the maximum $b-a \ge 0$.
If $b-a = 0$, the $\ZZ_2$-invariant of $Q$ is zero,
and the algorithm terminates.
Otherwise, we apply a ring homomorphism to every entry of $Q'$ where
every variable $x_j$ is mapped to a $(b-a) \times (b-a)$ matrix:
\begin{align}
	x_{j} 
	\mapsto 
	\begin{pmatrix}
	x_j &  & & \\
	    & x_j &  & \\
	    &     & \ddots &  \\
	    &     &        & x_j
	\end{pmatrix} \text{ for }j \neq d, 
	\qquad
	\text{ and } \quad
	x_d \mapsto \begin{pmatrix}
	0 & \cdots  & 0 & x_d \\
	1 &   &  & 0\\
	  & \ddots &  & \vdots \\
	  &   & 1& 0
	\end{pmatrix} \, . \label{eq:coarse-graining}
\end{align}
Note that the commutative invertible variables $x_j$ 
are mapped to commutative invertible matrices.
Every multiplication and addition of monomials
becomes that of matrices (ring homomorphism).
Overall, the $2q \times 2q$ matrix $Q'$ becomes a $2q(b-a) \times 2q(b-a)$ matrix $Q''$.
Since the $(b-a)$-th power of the image of $x_d$ is $x_d I_{b-a}$,
the new matrix $Q''$ has maximum exponent~$1$ for~$x_d$.
Now, write $Q''= A + x_d B$ where $A$ and $B$ do \emph{not} contain any $x_d$,
{\it i.e.},
both $A$ and $B$ have $x_d$-exponent zero and are matrices over 
$R_{d-1} = \FF_2[x_1^\pm,\ldots,x_{d-1}^\pm]$,
a Laurent polynomial ring over one fewer variables.

The matrix $B$ has $2q(b-a)$ columns,
but they are not always $R_{d-1}$-linearly independent.
Instead, the column span over $R_{d-1}$ has a free $R_{d-1}$-basis.
Suppose that such a free basis is written in the columns of a matrix $B'$.
This matrix has $2q(b-a)$ rows, 
but the number of columns may be anywhere from~$2$ to~$2q(b-a)$.
We finally form a matrix 
\begin{align}
	\Delta = (B')^\dag \lambda B'\, .
\end{align}
The matrix $\Delta$ is a nonsingular hermitian form in $d-1$ dimensions,
{\it i.e.}, $\det \Delta = 1 \in R_{d-1}$ and $\Delta^\dag = \Delta$.

\paragraph{Time reversal invariance and quadratic forms.}

If a translation-invariant Clifford QCA encoded in~$Q$ 
happens to map every real operator to a real operator,
then the matrix $Q^\dag \eta Q$, where
\begin{align}
	\eta = 
	\begin{pmatrix}
		0 & \one \\ 0 & 0
	\end{pmatrix},
\end{align}
satisfies a special property that its diagonal is free of scalars.
That is, $Q^\dag \eta Q = \eta + U + U^\dag$ for some matrix~$U$ over $R_d$.
(Recall that $1+1 = 0 \in R_d$.)
Since $\eta + \eta^\dag = \lambda$,
this condition automatically implies that $Q^\dag \lambda Q = \lambda$.
The construction of~$B'$ (whose columns form a free basis of the column span of $B$)
applies in this more special case,
and $\xi = (B')^\dag \eta B'$ is a nonsingular \emph{quadratic form},
{\it i.e.},
$\det(\xi +\xi^\dag) = 1$.
The hermitian form~$\Delta = \xi+\xi^\dag$ is called 
the associated hermitian form of $\xi$.

If $d=1$, then $\xi$ is over $\FF_2$.
The Witt group of all nonsingular quadratic forms over~$\FF_2$ 
is a group of order~$2$.
There are more efficient way to determine the $\ZZ_2$ invariant of quadratic forms,
but the following quantity (a special version of the Arf invariant) 
is convenient to state:
\begin{align}
	\Arf(\xi) = \frac{1}{2^{n/2}} \sum_{v \in \FF_2^n} (-1)^{v^T \xi v} = \pm 1
\end{align}
where $\xi$ is an $n\times n$ matrix.
Note that $\Arf(\xi \oplus \xi') = \Arf(\xi)\Arf(\xi')$.

\paragraph{From forms to time-reversal invariant QCA.}

Let $\Delta$ be a nonsingular \emph{hermitian} form over~$R_d$.
We need $\Delta^{-1}$, the matrix inverse.\footnote{%
While this must be easy,
we encountered some $\Delta = \Delta^\dag$ consisting of thousands of monomials
for which this inversion took many hours.
}
We ensure that every exponent of $x_d$ is one of $-1,0,1$ in both $\Delta$ and $\Delta^{-1}$.
To this end, as in the coarse-graining procedure for the map from QCA to forms,
we replace every $x_d$ by an invertible matrix of form in~\eqref{eq:coarse-graining}.
This replacement must be performed to $\Delta$ and $\Delta^{-1}$ simultaneously.
The size of the matrices blows up by some multiplicative factor.
The transformed matrices $\Delta$ and $\Delta^{-1}$ remain hermitian automatically.
Now, we have $\Delta = x_d^{-1} \Delta_1^\dag + \Delta_0 + x_d \Delta_1 = \Delta^\dag$
where $\Delta_0$ and $\Delta_1$ are free of~$x_d$.

Next, collect all the columns of two matrices
\begin{align}
	\Delta^{-1} (x_d^{-1} \Delta_1^\dag + \Delta_0 ) 
	\quad \text{ and }\quad
	\Delta^{-1} \Delta_1^\dag
\end{align}
to form a new matrix~$C$.
If $\Delta$ is $n \times n$,
then $C$ is an $n \times 2n$ matrix.%
\footnote{%
	This step amounts to finding a generating set for 
	a module $\partial_1(\Delta^{-1} B^*,B)$
	in the notation of~\cite[\S5.7]{haah2024topological}.
	This module is equal to 
	$\Delta^{-1} \projz^{\le 0} \Delta \projz^{\ge 0}$.  For a proof,
	observe that the projector $\Delta^{-1} \projz^{\le 0} \Delta$
	maps $\mathbf{z}^{\ge 0}$ into itself
	if we write $\projz^{\le 0} = \one - \projz^{\ge 1}$.
}
Construct a $2n \times 2n$ matrix $L$ by
\begin{align}
	L = \left[
		\begin{pmatrix} \projz^{\le 0} u \\ \projz^{\ge 0} \Delta u \end{pmatrix}
		~\middle|~
		u \text{ is a column of }C
		\right],
\end{align}
where $\projz^{\cdots}$ denotes a projection map 
that keeps only the monomials whose exponent in~$x_d$
is in the designated domain.
For example, $\projz^{\ge 0} (x_3 x_d^{-1} + x_1 x_2^{-1} x_d) = x_1 x_2^{-1} x_d$.
For matrices, apply $\projz^{\cdots}$ for all entries.
(The letter $z$ was used throughout \cite{haah2024topological} 
to refer to the last variable.)

Similarly, collect all the columns of the following two matrices
to form a new matrix~$D$,
\begin{align}
	\Delta^{-1}(\Delta_0 + x_d \Delta_1)\quad \text{ and }\quad \Delta^{-1} \Delta_1 \, ,
\end{align}
and construct a $2n \times 2n$ matrix $L^*$ by
\begin{align}
	L^* = \left[
		\begin{pmatrix} \projz^{\ge 0} g \\ \projz^{\ge 0} \Delta \projz^{<0} g \end{pmatrix}
		~\middle|~
		g \text{ is a column of }D
		\right]\, .
\end{align}
It turns out that $L$ and $L^*$ are both free of the variable $x_d$ 
and are both free modules over~$R_{d-1}$ of the same rank~$n$.
Find free bases for $L$ and $L^*$
and write them in one matrix $Q$
such that the first $n$ columns of $Q$ generate the column span of~$L$
and the last $n$ columns of~$Q$ generate that of~$L^*$.

The product $Q^\dagger \lambda Q$ will be 
of form $\begin{pmatrix} 0 & E \\ E^\dag & 0 \end{pmatrix}$
where $E$ is invertible.
The desired matrix that represents a Clifford QCA in $d-1$ dimensions 
is $Q ' = Q \, \diag(I, E^{-1})$.
It turns out that $Q'$ is time-reversal invariant.

So, the descent procedure starts with a Clifford QCA in $d$ dimensions,
giving a hermitian form over $d-1$ variables.
Then we get a time-reversal invariant Clifford QCA in dimension $d-2$,
and then a quadratic form over $d-3$ variables.
If we symmetrize it to get a hermitian form in $d-3$,
the procedure then enters a stage of stable recursion, all the way down to dimension zero.
If $d \ge 3$ is odd, then the final answer is given by the Arf invariant of a quadratic form 
over~$\FF_2$.
For even $d$, the procedure terminates with some Clifford QCA in zero dimensions,
which is always a circuit of finite depth 
and the original $d$-dimensional Clifford QCA
is always a shallow Clifford circuit plus a shift.


\subsection{Proving that for $k$ even, the $2k+1$ dimensional QCA is trivial} \label{5dtrivial}

Our $2k+1$ dimensional QCA is defined by the unitary
\begin{align*}
U^{WW}_{2k+1} = (-1)^{\int a \cup \rd b}
\end{align*}
We will show that we can write it as a circuit, using non-Clifford gates, when $k$ is even.  This means that for even $k$, the QCA $U^{WW}_{2k+1}$ are trivial.

Let us first give a heuristic argument for this in terms of the phases created by these unitaries.  As described in detail in \cite{fidkowski2023pumping} for the case $k=1$, the QCA $U^{WW}_{2k+1}$, when acting on a trivial product state, yields a higher dimensional Walker--Wang ground state.  This ground state is a superposition over two types of $k$-dimensional surfaces, and has a boundary state whose excitations contain two types of $(k-1)$-dimensional fermions.  In the case $k=1$ the bound state of these two now-pointlike fermions is again a fermion, and the result is the three fermion topological order, which is distinct from the toric code, which contains two bosons and a fermion.  This fact relies on the fact that the statistics of the bound state of two types of anyons with $\pi$ mutual braiding statistics is off by $\pi$ from the sum of their individual statistics.  This pattern persists for all odd $k$, but not, crucially, for even $k$.  For even $k$, the bound state of two fermions is a boson, so the resulting topological order is just the $(k,k)$ toric code \cite{Chen_2023}.  This topological order can be realized in $2k$ dimensions with commuting projectors, and hence indicates that the corresponding Walker--Wang model should be trivial.

To explicitly prove that the QCA $U^{WW}_{2k+1}$ is trivial, our first task will be to write a circuit which moves an $a$ charge around a $b$ flux.  Note that an $a$ charge is a $k-1$ dimensional object, which can be dragged across a $b$ flux, which is a $k$ dimensional object.  Specifically, we define:
\begin{align*}
U^{\text{braid}}&=\left(\prod_{\square_{k+1}}\left[ \left(\frac{1+\tG_{\square_{k+1}}}{2}\right)+ \left(\frac{1-\tG_{\square_{k+1}}}{2}\right)X_{\square_{k+1}} \right]\right)\\
&\cdot\left(\prod_{\square_{k+1}}\left[ \left(\frac{1+\tG_{\square_{k+1}}}{2}\right)+ \left(\frac{1-\tG_{\square_{k+1}}}{2}\right) \prod_{\square'_{k+1}} Z_{\square'_{k+1}}^{\int \square'_{k+1} \cup_1 \square_{k+1}}\right]\right)
\end{align*}
This is clearly a shallow depth circuit.  The gates that appear within it are ones that are controlled by $\tG_{\square_{k+1}}$, which in our notation is equal to $1-2\rd b(\square_{k+1})$.  Thus, on a configuration with a particular fixed value of $\rd b$, $U^{\text{braid}}$ acts
\begin{align*}
\left(\prod_{\square_{k+1} | \rd b(\square_{k+1})\neq 0} X_{\square_{k+1}}\right)\cdot \left(\prod_{\square'_{k+1}} Z_{\square'_{k+1}}^{\int \square'_{k+1} \cup_1 \rd b}\right)
\end{align*}
which is, up to sign, a product of short string operators around $\rd b$.  Note that the overall sign of such a product depends on the ordering of the string operators.  In the definition of $U^{\text{braid}}$, on the other hand, all of the $X$'s have been moved to the left of all of the $Z$'s.  

Let us now see how this acts on the states of our Hilbert space.  First, let us take the ground state of the first Walker--Wang model, with the second Walker--Wang model in a state corresponding to some flux configuration $\rd b$.  Such a ground state has amplitudes $(-1)^{\int \epsilon \cup \rd \epsilon}$ in the computational basis, as we derived above; namely $Z_{\square_{k+1}} = 1-2\epsilon(\square_{k+1})$.  Acting with $U^{\text{braid}}$ multiplies by $(-1)^{\int \rd \epsilon \cup_1 \rd b}$, coming from the $Z$ operators in $U^{\text{braid}}$, and shifts $\epsilon \rightarrow \epsilon+b$, coming from the $X$ operators in $U^{\text{braid}}$.  Thus the amplitude of $\epsilon'=\epsilon+b$ after this action is 
\begin{align*}
(-1)^{\int \epsilon \cup \rd \epsilon} (-1)^{\int \rd \epsilon \cup_1 \rd b} &= (-1)^{\int (\epsilon'+b) \cup d(\epsilon'+b)} (-1)^{\int d(\epsilon'+b) \cup_1 \rd b}\\
&= (-1)^{\int \epsilon' \cup \rd \epsilon'} (-1)^{\int b \cup \rd b}
\end{align*}
In other words, this state is an eigenstate of $U^{\text{braid}}$ with eigenvalue $(-1)^{\int b \cup \rd b}$.  Now let us consider a more general state, which corresponds to flux configurations $a$ and $b$.  Such a state can be created by acting with
\begin{align*}
\prod_{\square'_{k+1} | a(\hsquare_{k+1})\neq 0} Z_{\square_{k+1}}
\end{align*}
on the previous state.  The commutation of this past $U^{\text{braid}}$ introduces an Aharonov--Bohm braiding phase of $(-1)^{\int a\cup \rd b}$.  Now, since up to sign, for a given $\rd b$, $U^{\text{braid}}$ is a product of short string operators, it commutes with operators that create fermion excitations (the ones where the $\cup_1$ product appears in the opposite order).  It also commutes with operators that insert holonomies of the gauge field, since $\rd b$ is trivial in homology.  So $(-1)^{\int (b \cup \rd b + a \cup \rd b)}$ is the eigenvalue of $U^{\text{braid}}$ on any state with fixed $\rd b$ and $\rd a$.  Thus our QCA is equivalent, up to shallow circuits, to
\begin{align} \label{bcupdb}
U^{WW}_{2k+1} U^{\text{braid}} = (-1)^{\int b \cup \rd b}
\end{align}
Thus we just have to show that for $k$ even, the QCA defined by the unitary in \cref{bcupdb} is trivial.  We claim that in fact, in this case $\int b \cup \rd b$ can be written, mod $2$, as the integral of a local quantity that just depends on $\rd b$; this suffices.

First, we lift the $\Z_2$ valued $k$-chain $\beta$ to an integer valued $k$-chain $\tbeta$ arbitrarily.  Using eq. 29 in \cite{ChenTata}, we have:
\begin{align*}
d(\tbeta \cup_1 \tbeta)= d \tbeta \cup_1 d\tbeta + \tbeta \cup d\tbeta - d\tbeta \cup \tbeta
\end{align*}
Furthermore, for $k$ even, $d(\tbeta\cup\tbeta)=d\tbeta\cup\tbeta+\tbeta\cup d\tbeta$ (this is the crucial step where the fact that $k$ is even is used; for odd $k$ there is a relative minus sign between these terms).  Putting these two together and integrating, we obtain $0=\int d \tbeta \cup_1 d\tbeta + 2 \int \tbeta \cup d\tbeta$, or
\begin{align*}
\int \tbeta \cup d\tbeta = \frac{1}{2} \int d\tbeta \cup_1 d\tbeta.
\end{align*}
Note that, modulo $2$, $\int \tbeta \cup d\tbeta$ is equal to $\int \beta \cup d\beta$, so all we need to do is express $\frac{1}{2} \int d\tbeta \cup_1 d\tbeta$ as the integral of a local quantity.  Let $2\gamma = d\tbeta - \tdbeta$, where $\tdbeta$ is an arbitrary integer lift of $d\beta$ (notice that $d\tbeta - \tdbeta$ is even since its mod $2$ reduction clearly vanishes).  We now have, modulo $2$,
\begin{align*}
\int \tbeta \cup d\tbeta &= \frac{1}{2} \int \left(\tdbeta + 2\gamma\right) \cup_1 \left(\tdbeta + 2\gamma\right) \\
&= \frac{1}{2} \int \tdbeta \cup_1 \tdbeta + \int\left(\gamma \cup_1 \tdbeta + \tdbeta \cup_1 \gamma\right) + 2 \int \gamma \cup_1 \gamma \\
&= \frac{1}{2} \int \tdbeta \cup_1 \tdbeta + \int d\gamma \cup_2 \tdbeta \\
&= \frac{1}{2} \int \tdbeta \cup_1 \tdbeta + \frac{1}{2} \int d(d\tbeta-\tdbeta) \cup_2 \tdbeta \\
&= \frac{1}{2} \int \tdbeta \cup_1 \tdbeta - \frac{1}{2} \int d(\tdbeta) \cup_2 \tdbeta
\end{align*}
This is our local expression that can be evaluated by a shallow depth circuit, as it only depends on $d\beta$ rather than $\beta$.  Explicitly, we only care about the above expression modulo $2$, so we only care about $\tdbeta$ modulo $4$.  Note that both terms in the above expression can be expressed as a bounded sum of local terms, each of the form $\tdbeta(\square_{k+1}) \tdbeta(\square'_{k+1})$, where $\square_{k+1}$ and $\square'_{k+1}$ are some nearby $k+1$ cells.  This can be computed by a controlled gate on the two qubits residing on these two cells, with eigenvalue $+1$ unless $\tG(\square_{k+1}) = \tG(\square'_{k+1}) = -1$ in which case the eigenvalue is $i$ (owing to the overall $\frac{1}{2}$'s in the above expression).  Notice that this is a non-Clifford gate, consistent with the conjecture that this QCA is Clifford nontrivial.

\subsection{Clifford-hard, non-Clifford-easy}

The nontriviality or triviality of a QCA against quantum circuits
is always discussed with thermodynamic limits.
However, one can refine the question by measuring the complexity of a QCA.
An operator algebra automorphism of a finite dimensional simple complex algebra 
is always inner, realized by some unitary~$U$.
For a finite dimensional unitary $U$,
there always exists a local quantum circuit that implements $U$.
There can be various measures of the complexity of this local quantum circuit,
but since we consider bounded light cones
the most relevant complexity measure is the minimal \emph{circuit depth}
of a local quantum circuit where elementary gates couple neighboring sites.

Though we think of a QCA as one automorphism,
since our QCA~$\alpha$ is always translation invariant,
it actually gives an assignment of a finite unitary $U(\alpha,L)$
for each finite $d$-dimensional lattice of $L^d$ lattice points 
under periodic boundary conditions.
(It turns out that neither the translation invariance nor the periodic boundary condition 
is important; see~\cite{FHH}.)
The triviality of $\alpha$ means that
\begin{align}
	\lim_{L \to \infty}	\depth U(\alpha,L) < \infty .
\end{align}
When it comes to Clifford QCA, we can further refine the question
and restrict the set of local gates to be Clifford.
The complexity $\depth(U(\alpha,L);\mathrm{Clifford})$ with this restricted gate set
may be different from $\depth U(\alpha,L)$ without the restriction.

Indeed, 
if our conjecture that $U^{WW}_{2k+1}$ is Clifford nontrivial for $k \ge 1$ holds up,
then by the result above we have for $m \ge 1$
\begin{align}
	\lim_{L \to \infty}	\depth U^{WW}_{4m+1}(L) < \infty
	\quad\text{ and }\quad
	\lim_{L \to \infty}	\depth( U^{WW}_{4m+1}(L); \mathrm{Clifford} ) = \infty.
\end{align}
To the best of our knowledge,
this would be a first explicit example of a Clifford unitary 
on a finitely many qubits on a (Euclidean) metric space
that requires a deep circuit with local Clifford gates 
but a shallow circuit with non-Clifford gates.
We do not give a proof but it is conceivable that 
even if one allows elementary 2-site gates
to act on a pair of sites that are arbitrarily far apart
in a quantum circuit
({\it i.e.}, ignoring the complexity of swapping sites),
the divergence of the circuit depth as a function of the number of lattice sites
continues to hold.
Then, $U^{WW}_{4m+1}$ would be an example of a Clifford unitary
whose circuit depth (without the geometric locality constraint)
is high with Clifford gates but is low with non-Clifford gates.%
\footnote{There is a result of similar flavor~\cite{Ji2010}.}

\section{General QCA and Conjectured Stiefel--Whitney QCA Correspondence (SW-QCA)} \label{SWQCA}

Above, we have constructed a class of QCA $U^{WW}_{2k+1}$ in all odd spatial dimensions.  These QCA all square to the identity, and are time reversal invariant under a complex conjugation symmetry.  So, acting on a product state, they create an invertible state which is invariant under this time reversal\footnote{The resulting state is invertible, as two copies of the QCA is a quantum circuit since the QCA squares to the identity, so two copies of the resulting state is connected by a circuit to a product state.  Indeed, one may verify that using the time-reversal symmetry defined below, the state is still invertible as two copies of the QCA can be expressed as a time-reversal invariant circuit.}.  The question then arises, what is the field theory description of the invertible state that they create?  Here we conjecture that the action of this state is $\frac{1}{2}\int v_{k+1}^2$.  We also define a much broader class of QCA, and conjecture that their corresponding invertible phases are precisely the ones predicted by the cobordism classification of time reversal invariant SPT phases \cite{kapustin2014symmetry}.  We refer to this as the Stiefel--Whitney QCA (SW-QCA) correspondence.

To define these QCA, we consider a $d$-dimensional system with $m$ lattices, each slightly shifted with respect to the others. We choose non-negative integers $\{n_j\}$, $j=1,\ldots, m$ such that $\sum_{j=1}^m (n_j+1) = d+1$.  For each $j = 1,\ldots,m$ we also choose a sequence $\{i^j_s\}$, $s=1,\ldots,r_j$, with $\sum_{s=1}^{r_j} i^j_s = n_j+1$, $i^j_s \geq 2 i^j_{s+1}$ for $1\leq s \leq r_j-1$, and $i_{r_j} \leq d-n_j$.  On the $j$-th such lattice, we construct a Walker--Wang model Hamiltonian, whose corresponding action is ${\text{Sq}}^{i^j_1} \circ \ldots \circ {\text{Sq}}^{i^j_{r_j}} \left([e_j]\right)$, where $e$ is a $d-n_j$ form, as discussed in \cref{mostgen}.  Each such Walker-Wang model has charge and gauge flux excitations, with the gauge flux excitations being exact, i.e. boundaries of one higher dimensional manifolds.  We let $a_j$ be the $n_j$-forms Poincare dual to these manifolds; we will refer to these as the `gauge fields' corresponding to the gauge fluxes, which are Poincare dual to $\rd a_j$.  The QCA then acts by conjugation by a unitary that we denote $U^{a_1 \rd a_2 \ldots \rd a_m}$, which is defined as
\begin{align}
\label{genQCAunitary}
    U^{a_1 \rd a_2 \ldots \rd a_m}_{\bf{i}} \equiv (-1)^{\int a_1 \cup \rd a_2 \cup \ldots \cup \rd a_m}
\end{align}
Here ${\bf{i}}$ encodes the set of sequences $\{i^j_s\}$.  There are two special cases that will play an important role in the rest of the paper.  One is the case when $r_j=1$ for all $j$, so the sequences will just be $\{n_j+1\}$.  In that case we will denote $U^{a_1 \rd a_2 \ldots \rd a_m}_{\bf{i}}$ by $U^{a_1 \rd a_2 \ldots \rd a_m}_{\text{Wu}}$.  The other case is when all of the $i^j_s$ are powers of $2$; in that case we denote the corresponding QCA by $U^{a_1 \rd a_2 \ldots \rd a_m}_{\text{SW}}$.  In particular, the $2k+1$ dimensional Walker-Wang QCA $U^{WW}_{2k+1}$ of \cref{gen3FQCA} is $U^{WW}_{2k+1} = U^{a \rd b}_{\text{Wu}}$.

As before, $a_j$ themselves are not well defined local observables on the lattice, owing to the ambiguity in choosing a bounding manifold for the gauge fluxes.  However, the expression above may be interpreted as a mod $2$ linking of the gauge fluxes.  Namely, each gauge flux is a $d-n_j-1$ cycle dual to $\rd a_j$, and the cup product $\rd a_2 \cup \ldots \cup \rd a_m$ is dual to the intersection of the duals to $\rd a_2,\ldots,\rd a_m$, and we compute the mod $2$ linking of this intersection with the dual to $\rd a$.  This expression is symmetric under permuting the gauge fields, so it could equally be written $a_2 \cup \rd a_1 \cup \rd a_3 \cup \ldots \cup \rd a_m$, and so on.

Let us first consider the QCA $U^{a_1 \rd a_2 \ldots \rd a_m}_{\text{Wu}}$.  We will analyze the state produced by acting with $U^{a_1 \rd a_2 \ldots \rd a_m}_{\text{Wu}}$ on a certain product state, namely the state where the Walker--Wang models have no electric field lines / surfaces present (i.e., a product state in the $Z$ basis).  Our conjecture, supported by field theory computations, is that acting on this particular product state, the QCA $U^{a_1 \rd a_2 \ldots \rd a_m}_{\text{Wu}}$ produces the invertible state whose effective action on a general (possibly unorientable) spacetime manifold is given by a product of Wu classes:
\begin{align}
\label{WuAction}
 \int \prod_j v_{n_j+1},
\end{align}
where the product is a cup product.  We call this the {\emph{Wu-QCA correspondence}}, and we will give several arguments for it in \cref{attempt_field} and \cref{attempt_pumping}.  
  The action \cref{WuAction} is defined on unoriented manifolds, and is thus relevant for invertible phases protected by time-reversal symmetry \cite{kapustin2014symmetry}.  We conjecture that the appropriate action of time reversal symmetry here is a complex conjugation combined with a spin flip on Ising degrees of freedom, where the Ising degrees of freedom are those that describe electric `surfaces' in the $n_j=0$ cases (see \cref{nequals0}).  For notational simplicity, we will sometimes refer the QCA given by conjugation by $U^{a_1 \rd a_2 \ldots \rd a_m}_{\text{Wu}}$ as the ``$\prod_j v_{n_j+1}$ QCA", for the appropriate product.  For example, we will refer to the QCA $U^{WW}_{2k+1}$ of \cref{gen3FQCA} as the $v_{k+1}^2$ QCA.  In the special case of $k=1$, this is the $v_2^2=w_2^2+w_1^4$ QCA, which indeed produces the three fermion state when acting on the given product state \cite{fidkowski2023pumping}.

According to the cobordism classification, the most general effective action for SPTs protected by time reversal symmetry is not \cref{WuAction}, however, but rather an action given by an arbitrary product of Stiefel-Whitney classes.  The most general form of our conjecture, which we refer to as the \emph{Stiefel-Whitney-QCA (SW-QCA) correspondence}, is that any such general SPT can be produced by acting on a product state with some sequence of the QCAs $U^{a_1 \rd a_2 \ldots \rd a_m}_{\text{SW}}$.  In section \cref{attempt_field} we again substantiate this conjecture with a field theory argument, which includes an explicit algorithm for how to pick the sequence of QCAs $U^{a_1 \rd a_2 \ldots \rd a_m}_{\text{SW}}$, given some product of Stiefel-Whitney classes.


\subsection{Unitary $\Z_2$ symmetries}

Note that, in addition to being invariant under the standard complex conjugation time reversal symmetry, both the initial product state and the QCA are also invariant under unitary $\Z_2$ spin flip symmetries, one for each $j$ with $n_j=0$ (see \cref{nequals0}).  Thus we can probe the response of the invertible state that the QCA creates to background $\Z_2$ gauge fields $A_j$, for each $j$ such that $n_j=0$.  As an extension of the Wu-QCA correspondence, we conjecture that, for the QCA $U_{\text{Wu}}^{a_1 \rd a_2 \ldots \rd a_m}$,  the response theory for these unitary $\Z_2$ symmetries is given by the action
\begin{align}
\prod_{j, n_j>0} v_{n_j+1} \prod_{j,n_j=0} A_j,
\end{align}
meaning a cup product of Wu classes $v_{n_j+1}$ over $j$ such that $n_j>0$, times a cup product over gauge fields $A_j$.  In other words, we have replaced all instances of the first Wu class $v_1=w_1$ with the corresponding $A_j$.

For example, the QCA corresponding to $v_1^2=w_1^2$, namely $U^{a \, \rd a}_{\text{Wu}}$ with $a$ being a $0$-form, will just create the cluster state from the trivial state.  Let us derive this explicitly.  First, we only need a single instance of the $n=0$, $d=1$ Walker Wang model in this case, which consists of a single Ising degree of freedom.  The flux $\rd a$ is then represented by the plaquette term $Y_j Y_{j+1}$, as shown in \cref{nequals0}.  Thus the $0$-form $a$ itself can be represented by $Y_j$. To define $a \, \rd a$ we then need to shift $da$ to right, relative to $a$.  Then $\int a \, \rd a$ just counts the number of domain walls where we have a down spin to the left of the domain wall and an up spin to the right of the domain wall.  Here by `up' and `down' we mean $Y_j=1$ and $Y_j=-1$ respectively.  If we take periodic boundary conditions, then this is simply the number of total down spin domains.  Now let us take the trivial product state, stabilized by $Z_j = -1$ for all $j$.  After applying the QCA, the result will be stabilized by the conjugates of $Z_j$ by $(-1)^{\int a \, \rd a}$.  Now, $(-1)^{\int a \, \rd a} Z_j (-1)^{\int a \, \rd a}$ is simply an operator that flips the $j$th spin (again, in the $Y_j$ basis), and multiplies by $-1$ if and only if the total number of domain walls changes upon this flipping of the $j$th spin.  This happens if and only if $Y_{j-1} Y_{j+1}=1$.  In other words,
\begin{align*}
(-1)^{\int a \, \rd a} X_j (-1)^{\int a \, \rd a} = -Y_{j-1} Z_j Y_{j+1}
\end{align*}
The new state is thus stabilized by $\{ Y_{j-1} Z_j Y_{j+1} \}$.  This is the cluster state, which is the ground state of the $\Z_2 \times \Z_2$ SPT, in the same universality class as the Haldane phase.  Explicitly, the two $\Z_2$ symmetries are $\prod_{\text{even} \,j} Z_j$ and $\prod_{\text{odd} \,j} Z_j$, respectively.  However, in this context what is important is the overall $\Z_2$ symmetry $\prod_j Z_j$.  The cluster state is in particular also invariant under the composition of this $\Z_2$ symmetry and complex conjugation, which is how we defined our time reversal symmetry above.  This is just the standard action of time reversal on the cluster state, under which it is in the universality class of the Haldane chain, which has action $\int w_1^2$.  A similar calculation may be done to show that the QCA $U^{a_1 \, \rd a_2}$ for $0$-forms $a_1,a_2$ also creates a cluster state; this is similar to how it was found in \cite{fidkowski2023pumping} and more generally in \cref{5dtrivial} that the three-fermion QCA could be written either as a linking of two different gauge fields or as a self-linking of a single gauge field.

In general dimension $d$, again for a $0$-form $a$, we have the QCA $U_{\text{Wu}}^{a (\rd a)\cdots(\rd a)}$ which is degree $d+1$ in $a$.  For odd $d$ this creates the in-cohomology SPT state of time reversal, as defined in the previous paragraph, with action $w_1^{d+1}$, when acting on a trivial product state.  For even $d$, it creates the in-cohomology SPT state of the unitary $\Z_2$ symmetry.

\subsection{Field theory argument for the Wu-QCA and SW-QCA correspondences} \label{attempt_field}

Comparison of the action \cref{WuAction} with the QCA $U_{\text{Wu}}^{a_1 \rd a_2 \ldots \rd a_m}$ seems difficult at first: the action depends on a $d+1$-dimensional spacetime manifold and computes a scalar $\pm 1$, while the QCA acts on a $d$-dimensional space.  Likewise, the SW-QCA correspondence relating $U^{a_1 \rd a_2 \ldots \rd a_m}_{\text{SW}}$ defined in \cref{genQCAunitary} to general SPT actions formed from products of Stiefel-Whitney classes is at least as difficult.  Here we sketch some ideas which may be useful in proving this correspondence.

First, the hypercubic formalism used in constructing the QCAs in this paper is only applicable to flat space $\RR^{d}$.  In order to generalize to nontrivial spatial manifolds $M_d \times \RR$, we have to first extend our constructions to general simplicial manifolds.  Fortunately, the formalism of higher cup products does extend to such manifolds \cite{tata2020}, so we expect that our QCA constructions should also generalize.

First let us give a field theory `derivation' of these correspondences.  Let us consider a trivial tensor product state $|0\rangle$ in the Hilbert space on which $U^{a_1 \rd a_2 \ldots \rd a_m}_{\bf{i}}$ acts.  We view $|0\rangle$ as a resulting from the proliferation of gauge fluxes $\rd a_i$ on top of the ground states of the corresponding Walker-Wang models.  Explicitly, the ground state of the product of the Walker-Wang models is

\begin{align}
|{\text{g.s.}}\rangle = \int {\mathscr{D}} \epsilon_1\ldots {\mathscr{D}} \epsilon_m\, (-1)^{\sum_{j=1}^m \psi_j(\epsilon_j)} |\epsilon_1 \ldots \epsilon_m\rangle
\end{align}
where $\psi_j(\epsilon_j) \equiv \psi^{\{i^j_s\}}_{n_j,d}(\epsilon_j)$ is the wavefunction of the $j$th Walker-Wang model, given in particular in \cref{signgenp} in the case of trivial sequence $\{i^j_s\}$.  Note that the integral over the chains $\epsilon_j$ is really a discrete sum over $\Z_2$-valued cochains on a simplicial manifold.  Proliferating gauge fluxes on top of this state results in a trivial product state:

\begin{align}
|0\rangle = \int {\mathscr{D}} a_1\ldots {\mathscr{D}} a_m\,{\mathscr{D}} \epsilon_1\ldots {\mathscr{D}} \epsilon_m\, (-1)^{\sum_{j=1}^m \psi_j(\epsilon_j) + \int_{M_d} a_j \rd \epsilon_j} |\epsilon_1 \ldots \epsilon_m\rangle
\end{align}
Indeed, integrating out $a_j$ in the expression above forces $\rd \epsilon_j=0$, which just means that the result is a tensor product state where on each link the electric flux is trivial.  Acting on $|0\rangle$ by $U^{a_1 \rd a_2 \ldots \rd a_m}_{\bf{i}}$ results in

\begin{align}
U^{a_1 \rd a_2 \ldots \rd a_m}_{\bf{i}} |0\rangle = \int {\mathscr{D}} a_1\ldots {\mathscr{D}} a_m\,{\mathscr{D}} \epsilon_1\ldots {\mathscr{D}} \epsilon_m\, (-1)^{\sum_{j=1}^m \psi_j(\epsilon_j) + \int_{M_d} a_j \rd \epsilon_j + \int_{M_d} a_1 \rd a_2 \ldots \rd a_m} |\epsilon_1 \ldots \epsilon_m\rangle
\end{align}
This state represents an invertible phase whose partition function $\mathscr Z$ we can now attempt to compute on some spacetime manifold $M_{d+1}$.  Given that the bulk action corresponding to $\psi_j(\epsilon_j)$ is 

\begin{align*}
S_j(e_j)={\text{Sq}}^{i^j_1} \circ \ldots {\text{Sq}}^{i^j_{r_j}} \left(e_j\right)
\end{align*}
where the $d-n_j$ form $e_j$ corresponds to $d\epsilon_j$ at the boundary, it is natural to conjecture that

\begin{align}
{\mathscr Z}(M_{d+1}) = \int {\mathscr{D}} a_1\ldots {\mathscr{D}} a_m\,{\mathscr{D}} \epsilon_1\ldots {\mathscr{D}} \epsilon_m\, (-1)^{\int_{M_{d+1}} \left[\sum_{j=1}^m \left(S_j(\epsilon_j) + \rd a_j \,e_j\right) + \rd a_1 \rd a_2 \ldots \rd a_m \right]} 
\end{align}
where the path integrals are now over $d+1$ dimensional field configurations.  Indeed, this is just the natural action corresponding to the Hamiltonian that stabilizes the state $U^{a_1 \rd a_2 \ldots \rd a_m}_{\bf{i}} |0\rangle$.  Now
\begin{align}\label{alphadef}
{\text{Sq}}^{i^j_1} \circ \ldots {\text{Sq}}^{i^j_{r_j}} \left(e_j\right) = \alpha_{\{i^j_s\}} \cup e_j
\end{align}
for some cocycle $\alpha_{\{i^j_s\}}$ as shown in \cref{iteratedsquares} and \cite{mathoverflow2024}.  E.g. if $r_j=1$ then $\alpha_{\{i^j_s\}}$ is just the $(n_j+1)$th Wu class $v_{n_j+1}$.  Hence
\begin{align}
{\mathscr Z}(M_{d+1}) = \int {\mathscr{D}} a_1\ldots {\mathscr{D}} a_m\,{\mathscr{D}} \epsilon_1\ldots {\mathscr{D}} \epsilon_m\, (-1)^{\int_{M_{d+1}} \left[\sum_{j=1}^m \left(\alpha_{\{i^j_s\}} + \rd a_j\right)e_j \,+ \,\rd a_1 \rd a_2 \ldots \rd a_m \right]} 
\end{align}
Integrating out the $e_j$ then sets $\rd a_j=\alpha_{\{i^j_s\}}$, so the result is
\begin{align}
\label{bulkaction}
{\mathscr Z}(M_{d+1}) = (-1)^{\int_{M_{d+1}}  \alpha_{\{i^1_s\}}  \alpha_{\{i^2_s\}} \ldots \alpha_{\{i^m_s\}}}
\end{align}
which is the Wu-QCA correspondence in the case of trivial ${\bf{i}}$, i.e. in the case of 
$U^{a_1 \rd a_2 \ldots \rd a_m}_{\text{Wu}}$.

In the case when all the sequences $\{i^j_s\}$ consist of powers of $2$, i.e. in the case $U^{a_1 \rd a_2 \ldots \rd a_m}_{\text{SW}}$, as shown in \cref{iteratedsquares} we have
\begin{align}
\label{topSW}
\alpha_{\{i^j_s\}} = w_{n_j+1} + \ldots
\end{align}
 where the dots represent a polynomial in Stiefel-Whitney classes of degree strictly lower than $n_j+1$.
Assuming that \cref{bulkaction} correctly describes the action of the phase in this case,
it follows that we can create any action which is a sum of cup products of Stiefel-Whitney classes by composing various QCA of this form.
To see this, note that composition of QCA is the same, up to a circuit, as ``stacking" the QCA, i.e., taking a tensor product of several product states and acting with a different QCA on each one.  In this manner, we can produce an action which is a sum of actions of the form \cref{bulkaction} with $\alpha_{\{i^j_s\}} = w_{n_j+1} + \ldots$.  Thus, we have a linear map, with ${\mathbb F}_2$ coefficients, from QCA to actions.  Represent this linear map by a matrix, where each column corresponds to a QCA $U^{a_1 \rd a_2 \ldots \rd a_m}_{\text{SW}}$ and the corresponding row represents the action $\prod_j w_{n_j+1}$.  Then, \cref{topSW} implies that the diagonal elements of this matrix are equal to $1$.  Order the columns (and corresponding rows) so that any column with $m$ different gauge fields appears before any column with $m'$ gauge field if $m<m'$.  Then, this matrix is lower triangular as the $\ldots$ in \cref{topSW} increases the number of terms in the cup product.  Being a lower triangular square matrix with diagonal elements equal to $1$, it is invertible, proving that the map is onto.

What prevents this field theory argument from being a rigorous proof is the fact that there is a priori no universal way, given just the ground state wavefunction of an invertible state, to compute its partition function on arbitrary spacetime manifolds.  This is a problem common to all approaches that try to relate lattice Hamiltonian SPT models to spacetime effective actions.  In the following subsections we will give a complementary perspective on this construction by writing the QCA as being pumped out by a circuit in one higher dimension.  

\subsection{Pumping construction argument for the Wu-QCA and SW-QCA correspondences}
\label{attempt_pumping}

In order to make contact with invertible phases, which are diagnosed by evaluating the $U(1)$-valued partition function on topologically non-trivial Euclidean spacetime manifolds $M_{d+1}$, we must must find some way to associate a $U(1)$-valued scalar to a QCA and a manifold $M_{d+1}$.  To do this, recall  that for any
$d$-dimensional QCA, and any $d+1$-dimensional manifold with boundary, there is some quantum circuit that pumps
that QCA to the boundary via a ``swindle", while implementing the identity operation in the bulk\footnote{For any QCA $\alpha$, the tensor product $\alpha \otimes \alpha^{-1}$ is a quantum circuit.  We apply an Eilenberg swindle, using layers of such circuits so that everything cancels except $\alpha$ or $\alpha^{-1}$ at the boundary.  We learned this argument from A. Kitaev, but it is implicit in several papers.}.
So, on a $d+1$-dimensional manifold without boundary, such a pumping circuit will implement an identity, up to an overall complex scalar.  \emph{This is the scalar that we propose to associate with the given $d+1$-dimensional manifold.}

We emphasize that we are considering a spatial manifold in $d+1$-dimensions, rather than a spacetime manifold.
However, our proposal is not so different from proposals involving spacetime manifolds: given such a circuit on a $d+1$-dimensional spatial manifold, the trace of the corresponding unitary extracts the desired scalar, and this trace can be written as a tensor network in $d+1$ dimensions, similar to those used to define a spacetime path integral.

Of course, there is considerable freedom to choose different pumping circuits, so this complex scalar seems to be non-universal.  A similar phenomenon is known in the context of defining the action of an invertible phase on a spacetime lattice; see for example \cite{wanbranch} where a notion of ``branch independence" is proposed as a way to constrain the possible actions so that a universal scalar can be extracted.  We propose that 
\emph{given some $d$-dimensional QCA which implements conjugation by
$U^{a_1 \rd a_2 \ldots \rd a_m}_{\bf{i}}$, a circuit in $d+1$ dimensions which pumps this QCA and which obeys branch independence
is given by $U^{\rd a_1 \rd a_2 \ldots \rd a_m}_{\bf{i}}$.}  Here $U^{\rd a_1 \rd a_2 \ldots \rd a_m}_{\bf{i}}$ is defined analogously to $U^{a_1 \rd a_2 \ldots \rd a_m}_{\bf{i}}$.  In particular, the Hilbert space that it acts on is constructed in the same way as that of $U^{a_1 \rd a_2 \ldots \rd a_m}_{\bf{i}}$, namely as a tensor product of several Walker-Wang Hilbert spaces.  The only modification is that the dimension is increased from $d$ to $d+1$.  The ranks of the gauge fields $a_i$ are still $n_i$, and ${\bf{i}}$ remain unchanged.  The unitary $U^{\rd a_1 \rd a_2 \ldots \rd a_m}_{\bf{i}}$, however, is a circuit, since it computes the mod $2$ intersection number of the Poincare duals to the fluxes $\rd a_j$.  Since $\rd\left(a_1 \rd a_2 \ldots \rd a_m\right) = \rd a_1 \rd a_2 \ldots \rd a_m$, a simple Stokes' theorem argument shows that, on a manifold with boundary, this number is just equal to the integral of $\rd a_1 \rd a_2 \ldots \rd a_m$ over the boundary.

Given this circuit, we now try to compute the scalar given by $U^{\rd a_1 \rd a_2 \ldots \rd a_m}_{\bf{i}}$.  Let us focus on the case of trivial ${\bf{i}}$, which we denote by $U^{\rd a_1 \rd a_2 \ldots \rd a_m}_{\text{Wu}}$.  We claim that if the space $M_{d+1}$ has a nontrivial $v_k$, for $k>1$, then necessarily the flux of any $(k-1)$-form gauge field appearing in the corresponding Walker--Wang models used to define the QCA is nontrivial.  More precisely, the generalization of such Walker--Wang models to general triangulations and spatial manifolds of nontrivial topology should have the flux of the gauge field be in the same cohomology class as $v_k$.  
In the case of a $1$-form gauge field with $d=3$, this was explicitly shown previously \cite{chen2019bosonization} for a general triangulation with a branching structure on an orientable manifold\footnote{In that paper, the flux was computed to be in the same cohomology class as $w_2$ and $w_2=v_2$ on an orientable manifold.}.
Given this claim about flux, then the phase extracted by $U^{\rd a_1 \rd a_2 \ldots \rd a_m}_{\text{Wu}}$ is $-1$ raised to the mod $2$ intersection number of the Poincare duals of the gauge fluxes, which is now the same as the intersection number of the Poincare duals of the $v_k$, i.e. equal to $\int_{M_{d+1}} v_{n_1+1} \ldots v_{n_m+1}$.  This is what we wanted to show in the case of the Wu-QCA conjecture.  For the more general QCA, we replace the Wu class $v_{n_j+1}$ with the more general class $\alpha_{\{i_s^j\}}$ defined in \cref{alphadef} to obtain the general SW-QCA conjecture.

\subsubsection{Gauge Field Flux}
The claim that the gauge field flux and $v_k$ are in the same cohomology class can be derived explicitly in our lattice models.  We expect that this computation can be generalized to the case of $\alpha_{\{i_s^j\}}$, but for now we focus just on the case of $v_{n+1}$.  To compute the flux of the gauge field, consider some product of plaquette terms, where a plaquette term is a term in the second summation in \cref{HWWgen}.  In particular, we want to check if this product is $+1$ or $-1$ when taken over a set of plaquettes forming some cocycle $\epsilon$. 
 
 Each individual plaquette term is a product of some term $ G_{\square_{d-n-1}} $ from \cref{ndplaquette} multiplied by the 
 scalar  $(-1)^{\int \left(\square_{d-n-1} \cup_{d-2n-1} \rd\square_{d-n-1} + \square_{d-n-1} \cup_{d-2n-2} \square_{d-n-1}\right)}$.
 There is
 a natural guess for the product of these plaquettes: simply replace $\square_{d-n-1}$ with $\epsilon$ everywhere.
 That is, the guess is that the product is
 \begin{align}
 (-1)^{\int \left(\epsilon \cup_{d-2n-1} \rd\epsilon + \epsilon \cup_{d-2n-2} \epsilon\right)} G_{\epsilon},
 \end{align}
 where
 \begin{align}
 G_{\epsilon} &= 
    \prod_{\square_{d-n} \supset \epsilon} X_{\square_{d-n}} \left( \prod_{\square'_{d-n}} Z_{\square'_{d-n}} ^ {\int \rd \epsilon \cup_{d-2n} \square'_{d-n}} \right).
    \end{align}
    
 To prove this guess, iteratively change $\epsilon$, starting at $\epsilon=0$, going to some final $\epsilon$.  
 Suppose we add some $\square_{d-n-1}$ to $\epsilon$.
 Immediately below \cref{01sign} we computed the variation of the scalar $(-1)^{\int \left(\epsilon \cup_{d-2n-1} \rd\epsilon + \epsilon \cup_{d-2n-2} \epsilon\right)}$, giving $\epsilon$-dependent and -independent parts. 
 This $\epsilon$-independent part of the variation gives the scalar  $(-1)^{\int \left(\square_{d-n-1} \cup_{d-2n-1} \rd\square_{d-n-1} + \square_{d-n-1} \cup_{d-2n-2} \square_{d-n-1}\right)}$ in the definition of the plaquette term.
 The $\epsilon$-dependent part of the variation gives the extra term $\rd \square_{d-n-1} \cup_{d-2n} \rd\epsilon$.  However, note that $G_{\epsilon}$ is defined with all Pauli $Z$ operators ordered to the right of Pauli $X$ operators.
 So, when we left-multiply $G_{\epsilon}$ by some plaquette operator $G_{\square_{d-n-1}}$ to get $G_{\epsilon+\square_{d-n-1}}$ there is an extra scalar from commuting the $Z$ operators in $G_{\square_{d-n-1}}$ to the right of the $X$ operators in $G_{\epsilon}$.  This commutator term cancels the $\epsilon$-dependent part of the variation, verifying the guess.
 
Then, using this guess, if $\epsilon$ is a cocycle, the first term in the exponent of
$(-1)^{\int \left(\epsilon \cup_{d-2n-1} \rd\epsilon + \epsilon \cup_{d-2n-2} \epsilon\right)}$ vanishes, and the second term
in the exponent is the cup product $\epsilon \cup v_{n+1}$, and $G_{\epsilon}$ is equal to some product of Pauli $Z$ operators so it can be taken to be, say, $+1$ if evaluated in the $Z=+1$ state.

  \subsection{Ground State Properties of Invertible Phase}
To give more evidence for the SW-QCA and Wu-QCA correspondences.
we will consider properties of the ground state on spatial manifolds $M_d$ with nontrivial topology.  First let us focus on the Wu-QCA correspondence, and assume $M_d$ has non-trivial trivial $v_k$.  The argument here will only apply when the action does not contain $v_1$, but rather is made out of $v_k$ for $k>1$.
We will find that, for the given QCA, expectation values of certain observables are related to the Wu classes of the spatial manifold.  At the end of this section, we discuss to what extent these observables can be calculated for the given invertible phase using the action.
  
Now, consider some QCA $v_{n_1+1} v_{n_2+1} \ldots v_{n_m+1}$, and assume the space has nontrivial $v_{n_1+1} v_{n_2+1} \ldots v_{n_{m-1}+1}$.  Consider an operator $O$ which is a product of plaquette terms of the $m$-th gauge field over some set of plaquettes whose dual is a $(d-n_m)$-cycle $C$.   This operator $O$ could be thought of as a ``logical operator" of the gauge field, and such a product would be a logical operator if we considered an ordinary $\mathbb{Z}_2$ gauge field; however, here the gauge field has no logical qubits so the expectation value of $O$ has a definite sign.
Assume this cycle $C$ has nontrivial intersection with the dual to $v_{n_1+1} v_{n_2+1} \ldots v_{n_{m-1}+1}$.  We claim that in the state produced by acting with the QCA, this operator $O$ has expectation value $-1$.  For example, in the case of the $v_2^2$ theory, on a $3$-dimensional space with nontrivial $v_2$, such as $\mathbb{R}P^2 \times S^1$ for example, there is necessarily some nonvanishing flux; dually, this flux is a $1$-cycle.  The operator $O$ is a product of plaquettes over some set whose dual is a $2$-cycle, and we choose this $1$-cycle and this $2$-cycle to have nontrivial intersection.  To compute the expectation value of $O$, write it as a product of operators on individual $2$-cells, and the boundary of an odd number of those cells will link with the flux.

As another example, if we consider the $v_2^3$ theory on ${\mathbb{C}}P^2 \times S^1$, then this space has nontrivial $v_2^2$, so on this manifold, the expectation value of a certain operator $O$ changes sign.  This $v_2^3$ theory was considered previously \cite{fidkowski2023pumping}.  The flux of the $a_3$ gauge field is, dually, a $3$-cycle.  On this $3$-cycle, the QCA acts as $(-1)^{\int a_1 \rd a_2}$.
To understand the physics of this theory, take the third lattice (corresponding to gauge field $c$) much coarser than the first two lattices.  Then, the ground state is a superposition of $3$-cycles of the $a_3$ gauge field, with a
three-fermion Walker--Wang state of $a_1,a_2$ on each of those fluxes.

Can we calculate such observables using an action for an invertible phase described in terms of Stiefel--Whitney classes?
We do not know a good way to do this in general, though we can for specific examples, but if possible it would allow a more direct comparison of phases.  One approach to calculating them is as follows.
An observable such as $O$ can be thought of as introducing some flux for a trajectory in spacetime as follows.  The observable $O$ considered
above is the product of plaquettes dual to some cycle $(d-n_m)$-cycle $C$ in the spatial manifold.
Consider now some electrically charged $(n_m-1)$-spatial-dimensional excitation moving in spacetime.  The worldvolume of this excitation is $n_m$-dimensional.
If we imagine some worldvolume that has nontrivial intersection with $C$ for time $t<0$, avoids $C$ near $t=0$, and then again has nontrivial intersection for time $t>0$, then this worldvolume acquires a $-1$ phase due to the change in gauge field on $C$.
Thus, we can regard this as a flux inserted in spacetime, with the dual to the flux on a cycle $C \times \{0\}$, where the second coordinate in the product is the time coordinate.
As an example, imagine $0$-dimensional charged excitations moving in $\mathbb{C}P^2 \times S^1$, in the $\{1,f\}$ Walker--Wang model,
which corresponds to the $v_2^2$ QCA.
This spatial manifold has nontrivial $v_2$, and in fact $v_2$ has a representative which is dual to a point in $\mathbb{C}P^2$ times $S^1$.  Parameterize $S^1$ by $\mathbb{R}/\mathbb{Z}$.
Take $C$ to be dual to a homologically nontrivial $2$-cycle, such as $\mathbb{C}P^2 \times \{0\}$.
Consider a spacetime worldline for a $0$-dimensional charged excitation as follows.  This worldline is a $1$-cycle, and we take it to be the product of some point in $\mathbb{C}P^2$ with a $1$-cycle in
$S^1 \times \mathbb{R}$ where $\mathbb{R}$ is time.
We parameterize the cycle in $S^1 \times \mathbb{R}$ by a path that starts, for example, at coordinate $-1/2$ in $S^1$ at some $t<0$, moves positively in $S^1$ at fixed time until it reaches $+1/2$, then increases the time coordinate at $S^1$ coordinate until it reaches some $t>0$, then moves negatively in $S^1$ at fixed time until it reaches $-1/2$, the decreases the time coordinate at fixed $S^1$ coordinate until it returns to its start.  This path acquires a $-1$ phase.

So, we may consider whether there is some spacetime manifold that would necessarily induce such nontrivial flux in spacetime just as we picked $M$ to have some nontrivial Wu classes to induce flux.
Suppose we define a spacetime manifold that is a fiber bundle, with fiber $M$ and base $S^1$.  If the bundle is simply a product, $M \times S^1$,
then this describes a system on $M$, with periodic imaginary time on $S^1$.  We hope that by taking a nontrivial bundle, it may be
possible to choose the spacetime so that it has a nontrivial $v_{n_1+1} v_{n_2+1} \ldots v_{n_m+1}$.  

We conjecture that much of the above discussion can be generalized to the $U^{a_1 \rd a_2 \ldots \rd a_m}_{\text{SW}}$ QCA.  Then, we can at least partially implement this program for the $w_2 w_3$ theory.  Take spacetime to be the five-dimensional Dold manifold, which has nontrivial $w_2 w_3$ This manifold is a fiber bundle with fiber $\mathbb{C}P^2$ and base $S^1$.  It is given by $\mathbb{C}P^2 \times [0,1]/ (x \in \mathbb{C}P^2,0)\sim(\overline x,1)$ where the overline denotes complex conjugation.  The dual of $w_3$ is a nontrivial $2$-cycle, and can be represented by the nontrivial $2$-cycle in $\mathbb{C}P^2$ times a point in $S^1$.  We regard this as corresponding to such a logical operator $O$ at a given time.  This nontrivial $2$-cycle in $\mathbb{CP}^2$ is an $S^2$ and can be chosen invariant under complex conjugation.  The dual of $w_2$ is a nontrivial $3$-cycle, and can be represented by this $S^2$ in $\mathbb{C}P^2$ times $S^1$.  We regard this as corresponding to the flux of a $1$-form gauge field on spatial manifold $\mathbb{C}P^2$, which is present for all values of the ``time parameter" $S^1$.
 
\section{Future directions}

There are many potential future directions.  For one, several of the results in this work are still not rigorously proven.  The main outstanding issues are the generalization of QCA to branched triangulations and a full proof of the SW-QCA correspondence, showing that our $2k+1$ dimensional $U^{a\,\rd a}$ QCA is Clifford nontrivial, and a rigorous definition of braiding statistics for higher dimensional `fermionic' objects at the level of the lattice. 

Another issue is to understand better the various possible surface theories for the Walker--Wang states created by application of our higher dimensional QCA on a product state.  For the $3$-fermion QCA, an understanding of the anomalies underlying the corresponding surface state was key to proving that the QCA was nontrivial \cite{Haah_2022}.  It is possible that the same technique, coupled with an understanding of the surface states for the higher dimensional QCA, will allow one to prove that these QCA are nontrivial as well.  For the $2k+1$ dimensional $U^{a\,\rd a}_{\text{Wu}}$ QCA a natural guess is that it is nontrivial precisely for odd $k$, based on the discussion in \cref{5dtrivial}; it would be good to put these arguments on a firmer footing.

Another outstanding question is: when are the QCA arising from our construction non-trivial?  Certainly such a QCA is non-trivial if the corresponding invertible phase is absolutely stable, without any symmetry, as in, e.g., the $w_2 w_3$ case.  However, that does not exhaust the set of non-trivial QCA in our construction, since e.g. the $w_2^2$ QCA is non-trivial, whereas the corresponding phase is only protected by time reversal symmetry.  A natural conjecture, which is consistent with these and all other examples we know, is that a QCA arising from our construction is non-trivial precisely when the action of the corresponding SPT phase is non-trivial on some orientable spacetime manifold.  We leave the investigation of this conjecture, and the characterization of such SPT actions, to future work.

\begin{acknowledgments}
We thank Michael Freedman for useful conversations.  LF is supported by NSF DMR-2300172.
\end{acknowledgments}

\appendix

\section{Iterated Cup Products}
\label{iteratedsquares}

In \cref{Sq2Sq1}, we claimed that
$
{\text{Sq}}^2 {\text{Sq}}^1 \left( e\right) = (w_3 + w_1^2) \cup  e$.
To show this, we may use \cref{inductlemma} in this section, with $f(x)=x \cup v_2$, combined with the fact that
$\Sq^{2}(x_1)=x_1 \cup v_2$ for any $1$-cocycle $x_1$, with $v_j$ the $j$-th Wu class.  The claim then follows after some algebra, using $v_2=w_2+w_1^2$, as well as using the Wu formula for Steenrod squares of Stiefel-Whitney classes.

In this section we prove a more general result for iterated cup squares:
\begin{theorem}
\label{itsqthm}
Let $M$ be a smooth, $d$-dimensional, compact, connected manifold.  Let $x_{d-j}$ be a $\mathbb{Z}_2$-valued cocycle.
Let $k_1>k_2>\ldots>k_L$ be powers of $2$ such that $\sum_{a=1}^L k_a=j$.
Then,
\begin{align}
\label{itcupsquare}
\Sq^{k_1} \circ \Sq^{k_2} \circ \ldots \circ \Sq^{k_L}(x_{d-j})=x_{d-j} \cup y_j,
\end{align}
where
\begin{align}
y_j=w_j+\ldots,
\end{align}
where the $\ldots$ denotes a sum of products of Stiefel-Whitney classes, each of which is of degree {\emph smaller} than $j$.
\end{theorem}

Throughout the proof of the theorem, if we write that some cocycle is equal to $w_j+\ldots$, then the $\ldots$ will denote a sum of products of lower degree Stiefel-Whitney classes.
We will also need the following:
\begin{lemma}
\label{lucas}
Let $0\leq m<n$ be integers.  Let $m_i$ and $n_i$ denote the binary digits of $m,n$, respectively.  Then,
${m \choose n}=0 \mod 2 $ if and only if there is some $i$ such that $n_i=1$ and $m_i=0$.
\begin{proof}
This is Lucas's theorem in the case of mod $2$.
\end{proof}
\end{lemma}

We prove \cref{itsqthm} by induction on the number of iterated cup squares in \cref{itcupsquare}.

The base case is to show that
$\Sq^{j}(x_{d-j})=x_{d-j} \cup (w_j + \ldots)$.
Recall that $\Sq^{j}(x_{d-j})=x_{d-j} \cup v_j$, with $v_j$ the $j$-th Wu class.
So we need to show that $v_j=w_j+\ldots$ if $j$ is a power of $2$.
Recall that
the Stiefel-Whitney class $w$ is the Steenrod square of the Wu class $v$.   So, $w_j=v_j+\sum_{k<j} \Sq^{j-k} v_k$, or $v_j=w_j+\sum_{k<j} \Sq^{j-k} v_k$.
 
We show that for each $k<j$, the class $\Sq^{j-k} v_k$ does not contain $w_j$ in it, where we say it does not contain $w_j$ if it is a sum of products of lower degree Stiefel-Whitney classes.  Indeed, the only way it can contain $w_j$ is if $v_k$ includes $w_k$.   By the Wu formula, $\Sq^{j-k} w_k={j-k-1 \choose k} w_j+…$  However, ${j-k-1 \choose k}=0 \mod 2$.  To see this, consider the least significant nonvanishing binary digit of $k$.  The corresponding binary digit of $j-k-1$ vanishes and so the claim follows from \cref{lucas}. 

We now prove the inductive step.
We need the following:
\begin{lemma}
\label{inductlemma}
Let $f(x)$ be some operation on cocycles of some given degree so that $f(x)=x \cup y$ for some $y$.
Then,
\begin{align}
f \circ Sq^k(x)=x \cup z,
\end{align}
for
\begin{align}
\label{zeq}
z=y \cup v_k+\sum_{j=1}^k \sum_{j\leq i_1+\ldots+i_j\leq k} v_{k-i_1-\ldots-i_j} \cup  \Sq^{i_j} \circ \ldots \circ \Sq^{i_1}(y).
\end{align}
\begin{proof}
This proof is based on the answer at \cite{mathoverflow2024}.

By the Cartan formula,
$$\Sq^k(x)\cup y = \Sq^k(x \cup y)+\sum_{1 \leq i_1 \leq k} \Sq^{k-i_1}(x) \cup \Sq^{i_1}(y).$$

This formula iterates, and each term in the sum is
$$\Sq^{k-i_1}(x)\cup \Sq^{i_1}(y)=\Sq^{k-i_1}(x \cup \Sq^{i_1}(y)) + \sum_{1 \leq i_2 \leq b-i_1}\Sq^{k-i_1-i_2}(x) \cup \Sq^{i_2} \circ \Sq^{i_1}(y).$$

Thus,
$$\Sq^k(x)\cup y= \Sq^k(x \cup y) +\sum_{1 \leq i_1 \leq k} \Sq^{k-i_1}(x \cup \Sq^{i_1}(y)) + \sum_{2 \leq i_1+i_2 \leq k}
\Sq^{k-i_1-i_2}(x) \cup \Sq^{i_2} \circ \Sq^{i_1}(y).$$
Proceeding inductively, and using that $\Sq^{j}(x_{d-j})=x_{d-j} \cup v_j$ for any $j$, the lemma follows.
\end{proof}
\end{lemma}

Assuming the inductive hypothesis, we have
$\Sq^{k_1} \circ \Sq^{k_2} \circ \ldots \circ \Sq^{k_{L-1}}(x)=x\cup y,$
for $x$ of degree $d-k_1-\ldots-k_{L-1}$ and
for $y=w_{k_1+\ldots+k_{L-1}}+\ldots$.
So, by \cref{inductlemma},
we have
$\Sq^{k_1} \circ \Sq^{k_2} \circ \ldots \circ \Sq^{k_L}=x \cup z$
for $z$ in \cref{zeq}.
Each term in \cref{zeq} is a sum of cup products of Stiefel-Whitney classes of degree less than $k_1+\ldots+k_l$ except
for the second term in the case $i_1+\ldots+i_j=k_l$.
In that case, we use $y=w_{k_1+\ldots+k_{L-1}}+\ldots,$ so 
\begin{align}
z=\sum_{l=1}^{k_L} \sum_{l\leq i_1+\ldots+i_j=k_L} \Sq^{i_j} \circ \ldots \circ \Sq^{i_1}(w_{k_1+\ldots+k_{L-1}})+\ldots.
\end{align}
We can use the Wu formula to compute this sequence of Steenrod squares applied to a Stiefel-Whitney class.

Let $j=k_1+\ldots+k_L$.
One finds that
$z=c w_j+\ldots$
where the coefficient $c$
is
\begin{align}
\label{cdef}
c=\sum_{l=1}^k \sum_{j\leq i_1+\ldots+i_l=k_L} 
\prod_{s=1}^l {j-1+\sum_{t=1}^{s-1} i_t \choose i_s} \mod 2.
\end{align}

We claim that {\bf (1)} The term in \cref{cdef} with $l=1$ is equal to $1 \mod 2$ and {\bf (2)} for $l>1$, each term in the sum is equal to $0 \mod 2$.
This implies that $c=1$ and proves the inductive step.

To show {\bf (1)}:
\begin{lemma}
Let $k$ be a power of $2$ so that $k=2^m$ and let $j$ be a sum of powers of $2$, all of which are larger than $k$.  Then, ${j-1 \choose k}=1 \mod 2$.
\begin{proof}
This follows from \cref{lucas} since only one binary digit of $k$ is nonvanishing, and the corresponding binary digit of $j-1$ is equal to $1$.
\end{proof}
\end{lemma}

To show {\bf (2)}:
\begin{lemma}
Consider a sequence of integers $i_1,i_2,\ldots,i_l$ for some $l>1$, with $i_s\geq 1$ for all $s$ and $\sum_{s=1}^l i_s=k$ where $k$ is a power of $2$.  Let $j$ be a sum of powers of $2$, all of which are larger than $k$.
Then,
\begin{align}
\label{prodmod2}
\prod_{s=1}^l {j-1+\sum_{t=1}^{s-1} i_t \choose i_s}=0 \mod 2.
\end{align}
\begin{proof}
Consider the sequence $M_s$ defined by $M_s=j-1+\sum_{t=1}^{s-1} i_t$ for $t\in \{1,l+1\}$ so that
$M_1=j-1$ and $M_{l+1}=j-1+k$.
The first number $S_1$ in the sequence has its $m$ least significant binary digits all equal to $1$.  The last number $M_{l+1}$ also has its $m$ least significant binary digits equal to $1$.  Therefore, each of the $m$ least significant binary digits must change its value an \emph{even} number of times in this sequence, where we say a digit changes its value if its value differs between one element of the sequence $M_s$ and the next $M_{s+1}$.

Since $l>1$, we have $i_s<k$ for all $k$, so each $i_s$ has at least one nonvanishing binary digit among its $m$ least significant binary digits.
So, each $M_{s+1}$ differs from $M_s$ in at least one binary digit.
Consider the least significant binary digit such that for some $s$ that digit differs between $M_s$ and $M_{s+1}$.  This is the least significant binary digit that is nonvanishing in any $i_s$.
So, this digit is equal to $1$ for $M_1,M_2,\ldots,M_s$ and then is equal to $0$ for $M_{s+1}$.  It then remains equal to $0$ for $M_{s+1},M_{s+2},\ldots,M_{s'}$ before changing back to $1$ for some $M_{s'+1}$.
However, since this is the least significant binary digit that is nonvanishing in any $i_t$, if that digit changes from $0$ to $1$ then the corresponding binary digit in $i_{s'}$ equals $1$.  The product \cref{prodmod2} includes the term
${M_{s'} \choose i_{s'}}$, and the given binary digit in $s'$ equals $1$ while the given binary digit in $M_{s'}$ equals $0$ so this term is equal to $0 \mod 2$ by \cref{lucas}.
\end{proof}
\end{lemma}

This completes the proof of \cref{itsqthm}.

\bibliography{QCA}

\begin{thebibliography}{30}%
\makeatletter
\providecommand \@ifxundefined [1]{%
 \@ifx{#1\undefined}
}%
\providecommand \@ifnum [1]{%
 \ifnum #1\expandafter \@firstoftwo
 \else \expandafter \@secondoftwo
 \fi
}%
\providecommand \@ifx [1]{%
 \ifx #1\expandafter \@firstoftwo
 \else \expandafter \@secondoftwo
 \fi
}%
\providecommand \natexlab [1]{#1}%
\providecommand \enquote  [1]{``#1''}%
\providecommand \bibnamefont  [1]{#1}%
\providecommand \bibfnamefont [1]{#1}%
\providecommand \citenamefont [1]{#1}%
\providecommand \href@noop [0]{\@secondoftwo}%
\providecommand \href [0]{\begingroup \@sanitize@url \@href}%
\providecommand \@href[1]{\@@startlink{#1}\@@href}%
\providecommand \@@href[1]{\endgroup#1\@@endlink}%
\providecommand \@sanitize@url [0]{\catcode `\\12\catcode `\$12\catcode
  `\&12\catcode `\#12\catcode `\^12\catcode `\_12\catcode `\%12\relax}%
\providecommand \@@startlink[1]{}%
\providecommand \@@endlink[0]{}%
\providecommand \url  [0]{\begingroup\@sanitize@url \@url }%
\providecommand \@url [1]{\endgroup\@href {#1}{\urlprefix }}%
\providecommand \urlprefix  [0]{URL }%
\providecommand \Eprint [0]{\href }%
\providecommand \doibase [0]{http://dx.doi.org/}%
\providecommand \selectlanguage [0]{\@gobble}%
\providecommand \bibinfo  [0]{\@secondoftwo}%
\providecommand \bibfield  [0]{\@secondoftwo}%
\providecommand \translation [1]{[#1]}%
\providecommand \BibitemOpen [0]{}%
\providecommand \bibitemStop [0]{}%
\providecommand \bibitemNoStop [0]{.\EOS\space}%
\providecommand \EOS [0]{\spacefactor3000\relax}%
\providecommand \BibitemShut  [1]{\csname bibitem#1\endcsname}%
\let\auto@bib@innerbib\@empty
\bibitem [{\citenamefont {Haah}\ \emph {et~al.}(2023)\citenamefont {Haah},
  \citenamefont {Fidkowski},\ and\ \citenamefont {Hastings}}]{Haah_2022}%
  \BibitemOpen
  \bibfield  {author} {\bibinfo {author} {\bibfnamefont {Jeongwan}\
  \bibnamefont {Haah}}, \bibinfo {author} {\bibfnamefont {Lukasz}\ \bibnamefont
  {Fidkowski}}, \ and\ \bibinfo {author} {\bibfnamefont {Matthew~B.}\
  \bibnamefont {Hastings}},\ }\bibfield  {title} {\enquote {\bibinfo {title}
  {Nontrivial quantum cellular automata in higher dimensions},}\ }\href
  {\doibase 10.1007/s00220-022-04528-1} {\bibfield  {journal} {\bibinfo
  {journal} {Communications in Mathematical Physics}\ }\textbf {\bibinfo
  {volume} {398}},\ \bibinfo {pages} {469–540} (\bibinfo {year} {2023})},\
  \Eprint {http://arxiv.org/abs/1812.01625} {arXiv:1812.01625} \BibitemShut
  {NoStop}%
\bibitem [{\citenamefont {Gross}\ \emph {et~al.}(2012)\citenamefont {Gross},
  \citenamefont {Nesme}, \citenamefont {Vogts},\ and\ \citenamefont
  {Werner}}]{GNVW}%
  \BibitemOpen
  \bibfield  {author} {\bibinfo {author} {\bibfnamefont {D.}~\bibnamefont
  {Gross}}, \bibinfo {author} {\bibfnamefont {V.}~\bibnamefont {Nesme}},
  \bibinfo {author} {\bibfnamefont {H.}~\bibnamefont {Vogts}}, \ and\ \bibinfo
  {author} {\bibfnamefont {R.~F.}\ \bibnamefont {Werner}},\ }\bibfield  {title}
  {\enquote {\bibinfo {title} {Index theory of one dimensional quantum walks
  and cellular automata},}\ }\href {\doibase 10.1007/s00220-012-1423-1}
  {\bibfield  {journal} {\bibinfo  {journal} {Communications in Mathematical
  Physics}\ }\textbf {\bibinfo {volume} {310}},\ \bibinfo {pages} {419–454}
  (\bibinfo {year} {2012})},\ \Eprint {http://arxiv.org/abs/0910.3675}
  {arXiv:0910.3675} \BibitemShut {NoStop}%
\bibitem [{\citenamefont {Po}\ \emph {et~al.}(2016)\citenamefont {Po},
  \citenamefont {Fidkowski}, \citenamefont {Morimoto}, \citenamefont {Potter},\
  and\ \citenamefont {Vishwanath}}]{Po_2016}%
  \BibitemOpen
  \bibfield  {author} {\bibinfo {author} {\bibfnamefont {Hoi~Chun}\
  \bibnamefont {Po}}, \bibinfo {author} {\bibfnamefont {Lukasz}\ \bibnamefont
  {Fidkowski}}, \bibinfo {author} {\bibfnamefont {Takahiro}\ \bibnamefont
  {Morimoto}}, \bibinfo {author} {\bibfnamefont {Andrew~C.}\ \bibnamefont
  {Potter}}, \ and\ \bibinfo {author} {\bibfnamefont {Ashvin}\ \bibnamefont
  {Vishwanath}},\ }\bibfield  {title} {\enquote {\bibinfo {title} {Chiral
  {Floquet} phases of many-body localized bosons},}\ }\href {\doibase
  10.1103/physrevx.6.041070} {\bibfield  {journal} {\bibinfo  {journal}
  {Physical Review X}\ }\textbf {\bibinfo {volume} {6}},\ \bibinfo {pages}
  {041070} (\bibinfo {year} {2016})},\ \Eprint
  {http://arxiv.org/abs/1609.00006} {arXiv:1609.00006} \BibitemShut {NoStop}%
\bibitem [{\citenamefont {Po}\ \emph {et~al.}(2017)\citenamefont {Po},
  \citenamefont {Fidkowski}, \citenamefont {Vishwanath},\ and\ \citenamefont
  {Potter}}]{Po_2017}%
  \BibitemOpen
  \bibfield  {author} {\bibinfo {author} {\bibfnamefont {Hoi~Chun}\
  \bibnamefont {Po}}, \bibinfo {author} {\bibfnamefont {Lukasz}\ \bibnamefont
  {Fidkowski}}, \bibinfo {author} {\bibfnamefont {Ashvin}\ \bibnamefont
  {Vishwanath}}, \ and\ \bibinfo {author} {\bibfnamefont {Andrew~C.}\
  \bibnamefont {Potter}},\ }\bibfield  {title} {\enquote {\bibinfo {title}
  {Radical chiral floquet phases in a periodically driven {Kitaev} model and
  beyond},}\ }\href {\doibase 10.1103/physrevb.96.245116} {\bibfield  {journal}
  {\bibinfo  {journal} {Physical Review B}\ }\textbf {\bibinfo {volume} {96}},\
  \bibinfo {pages} {245116} (\bibinfo {year} {2017})},\ \Eprint
  {http://arxiv.org/abs/1701.01440} {arXiv:1701.01440} \BibitemShut {NoStop}%
\bibitem [{\citenamefont {Fidkowski}\ \emph {et~al.}(2019)\citenamefont
  {Fidkowski}, \citenamefont {Po}, \citenamefont {Potter},\ and\ \citenamefont
  {Vishwanath}}]{Fidkowski_2019}%
  \BibitemOpen
  \bibfield  {author} {\bibinfo {author} {\bibfnamefont {Lukasz}\ \bibnamefont
  {Fidkowski}}, \bibinfo {author} {\bibfnamefont {Hoi~Chun}\ \bibnamefont
  {Po}}, \bibinfo {author} {\bibfnamefont {Andrew~C.}\ \bibnamefont {Potter}},
  \ and\ \bibinfo {author} {\bibfnamefont {Ashvin}\ \bibnamefont
  {Vishwanath}},\ }\bibfield  {title} {\enquote {\bibinfo {title} {Interacting
  invariants for {Floquet} phases of fermions in two dimensions},}\ }\href
  {\doibase 10.1103/PhysRevB.99.085115} {\bibfield  {journal} {\bibinfo
  {journal} {Phys. Rev. B}\ }\textbf {\bibinfo {volume} {99}},\ \bibinfo
  {pages} {085115} (\bibinfo {year} {2019})},\ \Eprint
  {http://arxiv.org/abs/1703.12228} {arXiv:1703.12228} \BibitemShut {NoStop}%
\bibitem [{\citenamefont {Brun}\ and\ \citenamefont
  {Mlodinow}(2020)}]{Brun_2020}%
  \BibitemOpen
  \bibfield  {author} {\bibinfo {author} {\bibfnamefont {Todd~A.}\ \bibnamefont
  {Brun}}\ and\ \bibinfo {author} {\bibfnamefont {Leonard}\ \bibnamefont
  {Mlodinow}},\ }\bibfield  {title} {\enquote {\bibinfo {title} {Quantum
  cellular automata and quantum field theory in two spatial dimensions},}\
  }\href {\doibase 10.1103/PhysRevA.102.062222} {\bibfield  {journal} {\bibinfo
   {journal} {Phys. Rev. A}\ }\textbf {\bibinfo {volume} {102}},\ \bibinfo
  {pages} {062222} (\bibinfo {year} {2020})},\ \Eprint
  {http://arxiv.org/abs/2010.09104} {arXiv:2010.09104} \BibitemShut {NoStop}%
\bibitem [{\citenamefont {Fidkowski}\ \emph {et~al.}(2020)\citenamefont
  {Fidkowski}, \citenamefont {Haah},\ and\ \citenamefont {Hastings}}]{beyond}%
  \BibitemOpen
  \bibfield  {author} {\bibinfo {author} {\bibfnamefont {Lukasz}\ \bibnamefont
  {Fidkowski}}, \bibinfo {author} {\bibfnamefont {Jeongwan}\ \bibnamefont
  {Haah}}, \ and\ \bibinfo {author} {\bibfnamefont {Matthew~B.}\ \bibnamefont
  {Hastings}},\ }\bibfield  {title} {\enquote {\bibinfo {title} {Exactly
  solvable model for a $4+1\mathrm{D}$ beyond-cohomology symmetry-protected
  topological phase},}\ }\href {\doibase 10.1103/PhysRevB.101.155124}
  {\bibfield  {journal} {\bibinfo  {journal} {Phys. Rev. B}\ }\textbf {\bibinfo
  {volume} {101}},\ \bibinfo {pages} {155124} (\bibinfo {year} {2020})},\
  \Eprint {http://arxiv.org/abs/1912.05565} {arXiv:1912.05565} \BibitemShut
  {NoStop}%
\bibitem [{\citenamefont {Shirley}\ \emph {et~al.}(2022)\citenamefont
  {Shirley}, \citenamefont {Chen}, \citenamefont {Dua}, \citenamefont
  {Ellison}, \citenamefont {Tantivasadakarn},\ and\ \citenamefont
  {Williamson}}]{Shirley_2022}%
  \BibitemOpen
  \bibfield  {author} {\bibinfo {author} {\bibfnamefont {Wilbur}\ \bibnamefont
  {Shirley}}, \bibinfo {author} {\bibfnamefont {Yu-An}\ \bibnamefont {Chen}},
  \bibinfo {author} {\bibfnamefont {Arpit}\ \bibnamefont {Dua}}, \bibinfo
  {author} {\bibfnamefont {Tyler~D.}\ \bibnamefont {Ellison}}, \bibinfo
  {author} {\bibfnamefont {Nathanan}\ \bibnamefont {Tantivasadakarn}}, \ and\
  \bibinfo {author} {\bibfnamefont {Dominic~J.}\ \bibnamefont {Williamson}},\
  }\bibfield  {title} {\enquote {\bibinfo {title} {Three-dimensional quantum
  cellular automata from chiral semion surface topological order and beyond},}\
  }\href {\doibase 10.1103/prxquantum.3.030326} {\bibfield  {journal} {\bibinfo
   {journal} {PRX Quantum}\ }\textbf {\bibinfo {volume} {3}},\ \bibinfo {pages}
  {030326} (\bibinfo {year} {2022})},\ \Eprint
  {http://arxiv.org/abs/2202.05442} {arXiv:2202.05442} \BibitemShut {NoStop}%
\bibitem [{\citenamefont {Freedman}\ and\ \citenamefont
  {Hastings}(2020)}]{Freedman_2020}%
  \BibitemOpen
  \bibfield  {author} {\bibinfo {author} {\bibfnamefont {Michael}\ \bibnamefont
  {Freedman}}\ and\ \bibinfo {author} {\bibfnamefont {Matthew~B.}\ \bibnamefont
  {Hastings}},\ }\bibfield  {title} {\enquote {\bibinfo {title} {Classification
  of quantum cellular automata},}\ }\href {\doibase 10.1007/s00220-020-03735-y}
  {\bibfield  {journal} {\bibinfo  {journal} {Communications in Mathematical
  Physics}\ }\textbf {\bibinfo {volume} {376}},\ \bibinfo {pages} {1171–1222}
  (\bibinfo {year} {2020})},\ \Eprint {http://arxiv.org/abs/1902.10285}
  {arXiv:1902.10285} \BibitemShut {NoStop}%
\bibitem [{\citenamefont {Haah}(2021)}]{haah2021clifford}%
  \BibitemOpen
  \bibfield  {author} {\bibinfo {author} {\bibfnamefont {Jeongwan}\
  \bibnamefont {Haah}},\ }\bibfield  {title} {\enquote {\bibinfo {title}
  {Clifford quantum cellular automata: Trivial group in 2d and {Witt} group in
  3d},}\ }\href {\doibase 10.1063/5.0022185} {\bibfield  {journal} {\bibinfo
  {journal} {Journal of Mathematical Physics}\ }\textbf {\bibinfo {volume}
  {62}},\ \bibinfo {pages} {092202} (\bibinfo {year} {2021})},\ \Eprint
  {http://arxiv.org/abs/1907.02075} {arXiv:1907.02075} \BibitemShut {NoStop}%
\bibitem [{\citenamefont {Kapustin}(2014)}]{kapustin2014symmetry}%
  \BibitemOpen
  \bibfield  {author} {\bibinfo {author} {\bibfnamefont {Anton}\ \bibnamefont
  {Kapustin}},\ }\bibfield  {title} {\enquote {\bibinfo {title} {Symmetry
  protected topological phases, anomalies, and cobordisms: beyond group
  cohomology},}\ }\href@noop {} {\  (\bibinfo {year} {2014})},\ \Eprint
  {http://arxiv.org/abs/1403.1467} {arXiv:1403.1467} \BibitemShut {NoStop}%
\bibitem [{\citenamefont {Kitaev}(2006)}]{Kitaev_2006}%
  \BibitemOpen
  \bibfield  {author} {\bibinfo {author} {\bibfnamefont {Alexei}\ \bibnamefont
  {Kitaev}},\ }\bibfield  {title} {\enquote {\bibinfo {title} {Anyons in an
  exactly solved model and beyond},}\ }\href {\doibase
  10.1016/j.aop.2005.10.005} {\bibfield  {journal} {\bibinfo  {journal} {Annals
  of Physics}\ }\textbf {\bibinfo {volume} {321}},\ \bibinfo {pages} {2–111}
  (\bibinfo {year} {2006})},\ \Eprint {http://arxiv.org/abs/cond-mat/0506438}
  {arXiv:cond-mat/0506438} \BibitemShut {NoStop}%
\bibitem [{\citenamefont {Fidkowski}\ and\ \citenamefont
  {Hastings}(2023)}]{fidkowski2023pumping}%
  \BibitemOpen
  \bibfield  {author} {\bibinfo {author} {\bibfnamefont {Lukasz}\ \bibnamefont
  {Fidkowski}}\ and\ \bibinfo {author} {\bibfnamefont {Matthew~B}\ \bibnamefont
  {Hastings}},\ }\bibfield  {title} {\enquote {\bibinfo {title} {Pumping
  chirality in three dimensions},}\ }\href@noop {} {\  (\bibinfo {year}
  {2023})},\ \Eprint {http://arxiv.org/abs/2309.15903} {arXiv:2309.15903}
  \BibitemShut {NoStop}%
\bibitem [{\citenamefont {Chen}\ and\ \citenamefont
  {Kapustin}(2019)}]{chen2019bosonization}%
  \BibitemOpen
  \bibfield  {author} {\bibinfo {author} {\bibfnamefont {Yu-An}\ \bibnamefont
  {Chen}}\ and\ \bibinfo {author} {\bibfnamefont {Anton}\ \bibnamefont
  {Kapustin}},\ }\bibfield  {title} {\enquote {\bibinfo {title} {Bosonization
  in three spatial dimensions and a 2-form gauge theory},}\ }\href {\doibase
  10.1103/PhysRevB.100.245127} {\bibfield  {journal} {\bibinfo  {journal}
  {Physical Review B}\ }\textbf {\bibinfo {volume} {100}},\ \bibinfo {pages}
  {245127} (\bibinfo {year} {2019})},\ \Eprint
  {http://arxiv.org/abs/1807.07081} {arXiv:1807.07081} \BibitemShut {NoStop}%
\bibitem [{\citenamefont {Chen}\ and\ \citenamefont {Tata}(2023)}]{ChenTata}%
  \BibitemOpen
  \bibfield  {author} {\bibinfo {author} {\bibfnamefont {Yu-An}\ \bibnamefont
  {Chen}}\ and\ \bibinfo {author} {\bibfnamefont {Sri}\ \bibnamefont {Tata}},\
  }\bibfield  {title} {\enquote {\bibinfo {title} {Higher cup products on
  hypercubic lattices: application to lattice models of topological phases},}\
  }\href {\doibase 10.1063/5.0095189} {\bibfield  {journal} {\bibinfo
  {journal} {Journal of Mathematical Physics}\ }\textbf {\bibinfo {volume}
  {64}},\ \bibinfo {pages} {091902} (\bibinfo {year} {2023})},\ \Eprint
  {http://arxiv.org/abs/2106.05274} {arXiv:2106.05274} \BibitemShut {NoStop}%
\bibitem [{\citenamefont {Chen}(2020)}]{ChenPRR}%
  \BibitemOpen
  \bibfield  {author} {\bibinfo {author} {\bibfnamefont {Yu-An}\ \bibnamefont
  {Chen}},\ }\bibfield  {title} {\enquote {\bibinfo {title} {Exact bosonization
  in arbitrary dimensions},}\ }\href {\doibase
  10.1103/PhysRevResearch.2.033527} {\bibfield  {journal} {\bibinfo  {journal}
  {Phys. Rev. Res.}\ }\textbf {\bibinfo {volume} {2}},\ \bibinfo {pages}
  {033527} (\bibinfo {year} {2020})},\ \Eprint
  {http://arxiv.org/abs/1911.00017} {arXiv:1911.00017} \BibitemShut {NoStop}%
\bibitem [{\citenamefont {Fidkowski}\ \emph {et~al.}(2022)\citenamefont
  {Fidkowski}, \citenamefont {Haah},\ and\ \citenamefont
  {Hastings}}]{Fidkowski_gravitational}%
  \BibitemOpen
  \bibfield  {author} {\bibinfo {author} {\bibfnamefont {Lukasz}\ \bibnamefont
  {Fidkowski}}, \bibinfo {author} {\bibfnamefont {Jeongwan}\ \bibnamefont
  {Haah}}, \ and\ \bibinfo {author} {\bibfnamefont {Matthew~B.}\ \bibnamefont
  {Hastings}},\ }\bibfield  {title} {\enquote {\bibinfo {title} {Gravitational
  anomaly of $(3+1)$-dimensional ${\mathbb{z}}_{2}$ toric code with fermionic
  charges and fermionic loop self-statistics},}\ }\href {\doibase
  10.1103/PhysRevB.106.165135} {\bibfield  {journal} {\bibinfo  {journal}
  {Phys. Rev. B}\ }\textbf {\bibinfo {volume} {106}},\ \bibinfo {pages}
  {165135} (\bibinfo {year} {2022})},\ \Eprint
  {http://arxiv.org/abs/2110.14654} {arXiv:2110.14654} \BibitemShut {NoStop}%
\bibitem [{\citenamefont {Chen}\ and\ \citenamefont
  {Hsin}(2023)}]{Chen_2023anomaly}%
  \BibitemOpen
  \bibfield  {author} {\bibinfo {author} {\bibfnamefont {Yu-An}\ \bibnamefont
  {Chen}}\ and\ \bibinfo {author} {\bibfnamefont {Po-Shen}\ \bibnamefont
  {Hsin}},\ }\bibfield  {title} {\enquote {\bibinfo {title} {Exactly solvable
  lattice {Hamiltonians} and gravitational anomalies},}\ }\href {\doibase
  10.21468/scipostphys.14.5.089} {\bibfield  {journal} {\bibinfo  {journal}
  {SciPost Physics}\ }\textbf {\bibinfo {volume} {14}},\ \bibinfo {pages} {089}
  (\bibinfo {year} {2023})},\ \Eprint {http://arxiv.org/abs/2110.14644}
  {arXiv:2110.14644} \BibitemShut {NoStop}%
\bibitem [{\citenamefont {Haah}(2022)}]{haah2024topological}%
  \BibitemOpen
  \bibfield  {author} {\bibinfo {author} {\bibfnamefont {Jeongwan}\
  \bibnamefont {Haah}},\ }\href@noop {} {\enquote {\bibinfo {title}
  {Topological phases of unitary dynamics: Classification in {Clifford}
  category},}\ } (\bibinfo {year} {2022}),\ \Eprint
  {http://arxiv.org/abs/2205.09141} {arXiv:2205.09141} \BibitemShut {NoStop}%
\bibitem [{\citenamefont {Thorngren}(2018)}]{ThorngrenThesis}%
  \BibitemOpen
  \bibfield  {author} {\bibinfo {author} {\bibfnamefont {Ryan~George}\
  \bibnamefont {Thorngren}},\ }\emph {\bibinfo {title} {Combinatorial topology
  and applications to quantum field theory}},\ \href@noop {} {Ph.D. thesis},\
  \bibinfo  {school} {University of California, Berkeley} (\bibinfo {year}
  {2018})\BibitemShut {NoStop}%
\bibitem [{\citenamefont {Thorngren}(2015)}]{Thorngren_2015}%
  \BibitemOpen
  \bibfield  {author} {\bibinfo {author} {\bibfnamefont {Ryan}\ \bibnamefont
  {Thorngren}},\ }\bibfield  {title} {\enquote {\bibinfo {title} {Framed wilson
  operators, fermionic strings, and gravitational anomaly in 4d},}\ }\href
  {\doibase 10.1007/jhep02(2015)152} {\bibfield  {journal} {\bibinfo  {journal}
  {Journal of High Energy Physics}\ }\textbf {\bibinfo {volume} {2015}},\
  \bibinfo {pages} {152} (\bibinfo {year} {2015})},\ \Eprint
  {http://arxiv.org/abs/1404.4385} {arXiv:1404.4385} \BibitemShut {NoStop}%
\bibitem [{\citenamefont {Dub\'{e}}(1990)}]{Dube1990}%
  \BibitemOpen
  \bibfield  {author} {\bibinfo {author} {\bibfnamefont {Thomas~W.}\
  \bibnamefont {Dub\'{e}}},\ }\bibfield  {title} {\enquote {\bibinfo {title}
  {The structure of polynomial ideals and {Gröbner} bases},}\ }\href {\doibase
  10.1137/0219053} {\bibfield  {journal} {\bibinfo  {journal} {SIAM Journal on
  Computing}\ }\textbf {\bibinfo {volume} {19}},\ \bibinfo {pages} {750--773}
  (\bibinfo {year} {1990})}\BibitemShut {NoStop}%
\bibitem [{\citenamefont {Kera}\ \emph {et~al.}()\citenamefont {Kera},
  \citenamefont {Ishihara}, \citenamefont {Kambe}, \citenamefont {Vaccon},\
  and\ \citenamefont {Yokoyama}}]{Kera2023groebner}%
  \BibitemOpen
  \bibfield  {author} {\bibinfo {author} {\bibfnamefont {Hiroshi}\ \bibnamefont
  {Kera}}, \bibinfo {author} {\bibfnamefont {Yuki}\ \bibnamefont {Ishihara}},
  \bibinfo {author} {\bibfnamefont {Yuta}\ \bibnamefont {Kambe}}, \bibinfo
  {author} {\bibfnamefont {Tristan}\ \bibnamefont {Vaccon}}, \ and\ \bibinfo
  {author} {\bibfnamefont {Kazuhiro}\ \bibnamefont {Yokoyama}},\ }\href@noop {}
  {\enquote {\bibinfo {title} {Learning to compute gröbner bases},}\ }\Eprint
  {http://arxiv.org/abs/2311.12904} {arXiv:2311.12904} \BibitemShut {NoStop}%
\bibitem [{\citenamefont {Park}(1995)}]{Park1995}%
  \BibitemOpen
  \bibfield  {author} {\bibinfo {author} {\bibfnamefont {Hyung-Ju}\
  \bibnamefont {Park}},\ }\emph {\bibinfo {title} {A computational theory of
  {Laurent} polynomial rings and multidimensional {FIR} systems}},\ \href
  {https://www2.eecs.berkeley.edu/Pubs/TechRpts/1995/ERL-95-39.pdf} {Ph.D.
  thesis},\ \bibinfo  {school} {University of California, Berkeley} (\bibinfo
  {year} {1995})\BibitemShut {NoStop}%
\bibitem [{\citenamefont {Chen}\ \emph {et~al.}(2023)\citenamefont {Chen},
  \citenamefont {Dua}, \citenamefont {Hsin}, \citenamefont {Jian},
  \citenamefont {Shirley},\ and\ \citenamefont {Xu}}]{Chen_2023}%
  \BibitemOpen
  \bibfield  {author} {\bibinfo {author} {\bibfnamefont {Xie}\ \bibnamefont
  {Chen}}, \bibinfo {author} {\bibfnamefont {Arpit}\ \bibnamefont {Dua}},
  \bibinfo {author} {\bibfnamefont {Po-Shen}\ \bibnamefont {Hsin}}, \bibinfo
  {author} {\bibfnamefont {Chao-Ming}\ \bibnamefont {Jian}}, \bibinfo {author}
  {\bibfnamefont {Wilbur}\ \bibnamefont {Shirley}}, \ and\ \bibinfo {author}
  {\bibfnamefont {Cenke}\ \bibnamefont {Xu}},\ }\bibfield  {title} {\enquote
  {\bibinfo {title} {Loops in 4+1d topological phases},}\ }\href {\doibase
  10.21468/scipostphys.15.1.001} {\bibfield  {journal} {\bibinfo  {journal}
  {SciPost Physics}\ }\textbf {\bibinfo {volume} {15}},\ \bibinfo {pages} {001}
  (\bibinfo {year} {2023})},\ \Eprint {http://arxiv.org/abs/2112.02137}
  {arXiv:2112.02137} \BibitemShut {NoStop}%
\bibitem [{\citenamefont {Freedman}\ \emph {et~al.}(2022)\citenamefont
  {Freedman}, \citenamefont {Haah},\ and\ \citenamefont {Hastings}}]{FHH}%
  \BibitemOpen
  \bibfield  {author} {\bibinfo {author} {\bibfnamefont {Michael}\ \bibnamefont
  {Freedman}}, \bibinfo {author} {\bibfnamefont {Jeongwan}\ \bibnamefont
  {Haah}}, \ and\ \bibinfo {author} {\bibfnamefont {Matthew~B.}\ \bibnamefont
  {Hastings}},\ }\bibfield  {title} {\enquote {\bibinfo {title} {The group
  structure of quantum cellular automata},}\ }\href {\doibase
  10.1007/s00220-022-04316-x} {\bibfield  {journal} {\bibinfo  {journal}
  {Commun. Math. Phys.}\ }\textbf {\bibinfo {volume} {389}},\ \bibinfo {pages}
  {1277--1302} (\bibinfo {year} {2022})},\ \Eprint
  {http://arxiv.org/abs/1910.07998} {arXiv:1910.07998} \BibitemShut {NoStop}%
\bibitem [{\citenamefont {Ji}\ \emph {et~al.}(2010)\citenamefont {Ji},
  \citenamefont {Chen}, \citenamefont {Wei},\ and\ \citenamefont
  {Ying}}]{Ji2010}%
  \BibitemOpen
  \bibfield  {author} {\bibinfo {author} {\bibfnamefont {Zhengfeng}\
  \bibnamefont {Ji}}, \bibinfo {author} {\bibfnamefont {Jianxin}\ \bibnamefont
  {Chen}}, \bibinfo {author} {\bibfnamefont {Zhaohui}\ \bibnamefont {Wei}}, \
  and\ \bibinfo {author} {\bibfnamefont {Mingsheng}\ \bibnamefont {Ying}},\
  }\bibfield  {title} {\enquote {\bibinfo {title} {The {LU-LC} conjecture is
  false},}\ }\href@noop {} {\bibfield  {journal} {\bibinfo  {journal} {Quantum
  Inf. Comput.}\ }\textbf {\bibinfo {volume} {10}},\ \bibinfo {pages} {97--108}
  (\bibinfo {year} {2010})},\ \Eprint {http://arxiv.org/abs/0709.1266}
  {arXiv:0709.1266} \BibitemShut {NoStop}%
\bibitem [{\citenamefont {Tata}(2020)}]{tata2020}%
  \BibitemOpen
  \bibfield  {author} {\bibinfo {author} {\bibfnamefont {Sri}\ \bibnamefont
  {Tata}},\ }\bibfield  {title} {\enquote {\bibinfo {title} {Geometrically
  interpreting higher cup products, and application to combinatorial pin
  structures},}\ }\href@noop {} {\  (\bibinfo {year} {2020})},\ \Eprint
  {http://arxiv.org/abs/2008.10170} {arXiv:2008.10170} \BibitemShut {NoStop}%
\bibitem [{mat()}]{mathoverflow2024}%
  \BibitemOpen
  \href@noop {} {}\bibinfo {note}
  {\url{https://mathoverflow.net/questions/481747/formula-for-compositions-of-steenrod-squares-that-produce-a-form-in-the-top-degr}}\BibitemShut
  {NoStop}%
\bibitem [{\citenamefont {Wan}\ \emph {et~al.}(2022)\citenamefont {Wan},
  \citenamefont {Wang},\ and\ \citenamefont {Wen}}]{wanbranch}%
  \BibitemOpen
  \bibfield  {author} {\bibinfo {author} {\bibfnamefont {Zheyan}\ \bibnamefont
  {Wan}}, \bibinfo {author} {\bibfnamefont {Juven}\ \bibnamefont {Wang}}, \
  and\ \bibinfo {author} {\bibfnamefont {Xiao-Gang}\ \bibnamefont {Wen}},\
  }\bibfield  {title} {\enquote {\bibinfo {title} {(3+ 1) d boundaries with
  gravitational anomaly of (4+ 1) d invertible topological order for
  branch-independent bosonic systems},}\ }\href@noop {} {\bibfield  {journal}
  {\bibinfo  {journal} {Physical Review B}\ }\textbf {\bibinfo {volume}
  {106}},\ \bibinfo {pages} {045127} (\bibinfo {year} {2022})}\BibitemShut
  {NoStop}%
\end{thebibliography}%
\end{document}